\documentclass[longauth]{aa} 

\usepackage{graphicx}
\usepackage{txfonts}
\usepackage[pdfpagelabels=false]{hyperref}	
\hypersetup{colorlinks=true,linkcolor=blue,citecolor=blue,filecolor=blue,urlcolor=blue,}

\newcommand*{\orcid}[1]{
    \href{https://orcid.org/#1}{\,\raisebox{0.2em}{
    \includegraphics[height=0.6em,width=0.6em]{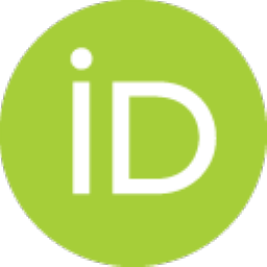}
}}}

\usepackage{multirow}  
\usepackage{xcolor} 

\usepackage{xspace}     

\newcommand*{\hcnone}{\ensuremath{\text{HCN(1--0)}}\xspace} 
\newcommand*{\hcopone}{\ensuremath{\text{HCO}^+\text{(1--0)}}\xspace} 
\newcommand*{\cstwo}{\ensuremath{\text{CS(2--1)}}\xspace} 
\newcommand*{\coone}{\ensuremath{\text{CO(1--0)}}\xspace} 
\newcommand*{\cotwo}{\ensuremath{\text{CO(2--1)}}\xspace} 
\newcommand*{\htwo}{\ensuremath{\text{H}_2}\xspace} 
\newcommand*{\hone}{\ensuremath{\text{H}\,{\tiny\text{I}}}\xspace}  
\newcommand*{\halpha}{\ensuremath{\text{H}\alpha}\xspace} 
\newcommand*{\nnhp}{\ensuremath{\text{N}_2\text{H}^+}\xspace} 

\newcommand*{\intCO}{\ensuremath{W_{\text{CO(2--1)}}}\xspace}  
\newcommand*{\intHCN}{\ensuremath{W_{\text{HCN}}}\xspace}  
\newcommand*{\sigmol}{\ensuremath{\Sigma_{\text{mol}}}\xspace}  
\newcommand*{\sigmolcloud}{\ensuremath{\Sigma_{\text{mol, 120\,pc}}}\xspace}  
\newcommand*{\sigmollarge}{\ensuremath{\Sigma_{\text{mol, 260\,pc}}}\xspace}  
\newcommand*{\vdis}{\ensuremath{\sigma_{\text{mol}}}\xspace}  
\newcommand*{\avir}{\ensuremath{\alpha_{\text{vir}}}\xspace}  
\newcommand*{\PDE}{\ensuremath{P_{\text{DE}}}\xspace}  
\newcommand*{\Pcloud}{\ensuremath{P_{\text{cloud}}}\xspace}  
\newcommand*{\Patom}{\ensuremath{P_{\text{atom}}}\xspace}  
\newcommand*{\aCO}{\ensuremath{\alpha_{\text{CO}}}\xspace}  
\newcommand*{\aHCN}{\ensuremath{\alpha_{\text{HCN}}}\xspace}  
\newcommand*{\neff}{\ensuremath{n_{\text{eff}}}\xspace}  
\newcommand*{\sigsfr}{\ensuremath{\Sigma_{\text{SFR}}}\xspace}  
\newcommand*{\sfrha}{\ensuremath{\text{SFR}_{\mathrm{H}\alpha}}\xspace}  
\newcommand*{\mmol}{\ensuremath{M_{\text{mol}}}\xspace}  
\newcommand*{\mdense}{\ensuremath{M_{\text{dense}}}\xspace}  
\newcommand*{\sigdense}{\ensuremath{\Sigma_{\text{dense}}}\xspace}  
\newcommand*{\fdense}{\ensuremath{f_{\text{dense}}}\xspace}  
\newcommand*{\sfedense}{\ensuremath{\text{SFE}_{\text{dense}}}\xspace}  
\newcommand*{\sfemol}{\ensuremath{\text{SFE}_{\text{mol}}}\xspace}  
\newcommand*{\sigstar}{\ensuremath{\Sigma_{\star}}\xspace}  
\newcommand*{\rhostar}{\ensuremath{\rho_{\star}}\xspace}  
\newcommand*{\sigatom}{\ensuremath{\Sigma_{\text{atom}}}\xspace}  
\newcommand*{\siggas}{\ensuremath{\Sigma_{\text{gas}}}\xspace}  

\newcommand*{\sigmolavg}{\ensuremath{\langle\Sigma_{\text{mol}}\rangle}\xspace}
\newcommand*{\vdisavg}{\ensuremath{\langle\sigma_{\text{mol}}\rangle}\xspace}
\newcommand*{\aviravg}{\ensuremath{\langle\alpha_{\text{vir}}\rangle}\xspace}

\newcommand*{\PDEavg}{\ensuremath{\langle P_{\text{DE}}\rangle}\xspace}
\newcommand*{\Pcloudavg}{\ensuremath{\langle P_{\text{cloud}}\rangle}\xspace}

\newcommand*{\rgal}{\ensuremath{r_{\rm gal}}\xspace} 
\newcommand*{\gal}{\ensuremath{\mathrm{NGC\,4321}}\xspace}  

\usepackage{siunitx}
\usepackage{xparse}  
\DeclareSIUnit \parsec {pc}  
\DeclareSIUnit \micron {\micro\metre}  
\DeclareSIUnit \year {yr}  
\DeclareSIUnit \jansky {Jy}  
\DeclareSIUnit \Msun {M_{\odot}}  
\DeclareSIUnit \Lsun {L_{\odot}}  
\DeclareSIUnit \Kkms {\kelvin\km\per\second}  
\DeclareSIUnit \kB {\textit{k}_B}  
\DeclareSIUnit \dex {dex}  
\DeclareSIUnit \erg {erg}  
\NewDocumentCommand\angRange{O{} m m}{\SIrange[parse-numbers=false, #1]{\ang[parse-numbers=true]{#2}}{\ang[parse-numbers=true]{#3}}{}}  
\newcommand*{\ra}[2][]{{
    \def\SIUnitSymbolDegree{\textsuperscript{h}}%
    \def\SIUnitSymbolArcminute{\textsuperscript{m}}%
    \def\SIUnitSymbolArcsecond{\textsuperscript{s}}%
    \ang[#1]{#2}}%
}

\begin{document} 

    \title{A 260 pc resolution ALMA map of HCN(1--0) in the galaxy NGC\,4321}
    
    \titlerunning{HCN and star formation at 260\,pc in NGC\,4321}


   \author{Lukas~Neumann
          \inst{1}\fnmsep\thanks{\email{lukas.neumann.astro@gmail.com}}
          \thanks{Member of the International Max Planck Research School (IMPRS) for Astronomy and Astrophysics at the Universities of Bonn and Cologne.}
          \orcid{0000-0001-9793-6400}
          \and
          Frank~Bigiel\inst{1}\orcid{0000-0003-0166-9745}
          \and
          Ashley~T.~Barnes\inst{2}\orcid{0000-0003-0410-4504}
          \and
          Molly J.~Gallagher\inst{3}\orcid{0000-0001-5285-5930}
          \and
          Adam~Leroy\inst{3}\orcid{0000-0002-2545-1700}
          \and
          Antonio~Usero\inst{4}\orcid{0000-0003-1242-505X}
          \and
          Erik~Rosolowsky\inst{5}\orcid{0000-0002-5204-2259}
          \and
          Ivana~Be\v{s}li\'c\inst{6}\orcid{0000-0003-0583-7363}
          \and
          Médéric~Boquien\inst{7}\orcid{0000-0003-0946-6176}
          \and
          Yixian~Cao\inst{8}\orcid{0000-0001-5301-1326}
          \and
          M\'elanie~Chevance\inst{9,10}\orcid{0000-0002-5635-5180}
          \and
          Dario~Colombo\inst{1}\orcid{0000-0001-6498-2945}
          \and
          Daniel~A.~Dale\inst{11}\orcid{0000-0002-5782-9093}
          \and
          Cosima~Eibensteiner\inst{1,12}\orcid{0000-0002-1185-2810}
          \and
          Kathryn~Grasha\inst{13,14,15}\orcid{0000-0002-3247-5321}
          \and
          Jonathan~D.~Henshaw\inst{16,17}\orcid{0000-0001-9656-7682}
          \and
          María~J.~Jiménez-Donaire\inst{4,18}\orcid{0000-0002-9165-8080}
          \and
          Sharon~Meidt\inst{19}\orcid{0000-0002-6118-4048}
          \and
          Shyam H.~Menon\inst{20}\orcid{0000-0001-5944-291X}
          \and
          Eric~J.~Murphy\inst{12}\orcid{0000-0001-7089-7325}
          \and
          Hsi-An~Pan\inst{21}\orcid{0000-0002-1370-6964}
          \and
          Miguel~Querejeta\inst{4}\orcid{0000-0002-0472-1011}
          \and
          Toshiki~Saito\inst{22}\orcid{0000-0002-2501-9328}
          \and
          Eva~Schinnerer\inst{17}\orcid{0000-0002-3933-7677}
          \and
          Sophia~K.~Stuber\inst{17}\orcid{0000-0002-9333-387X}
          \and
          Yu-Hsuan~Teng\inst{23}\orcid{0000-0003-4209-1599}
          \and
          Thomas~G.~Williams\inst{24}\orcid{0000-0002-0012-2142}
          }

   \institute{
    Argelander-Institut für Astronomie, Universität Bonn, Auf dem Hügel 71, 53121 Bonn, Germany
    \and
    European Southern Observatory, Karl-Schwarzschild Stra{\ss}e 2, D-85748 Garching bei M\"{u}nchen, Germany
    \and    
    Department of Astronomy, The Ohio State University, 140 West 18th Ave, Columbus, OH 43210, USA
    \and    
    Observatorio Astron\'omico Nacional (IGN), C/ Alfonso XII, 3, E-28014 Madrid, Spain
    \and    
    Dept. of Physics, University of Alberta, Edmonton, Alberta, Canada T6G 2E1
    \and
    LERMA, Observatoire de Paris, PSL Research University, CNRS, Sorbonne Universit\'es, 75014 Paris, France
    \and
    Université Côte d'Azur, Observatoire de la Côte d'Azur, CNRS, Laboratoire Lagrange, 06000, Nice, France
    \and
    Max-Planck-Institut f\"ur Extraterrestrische Physik (MPE), Giessenbachstr. 1, D-85748 Garching, Germany
    \and
    Instit\"ut  f\"{u}r Theoretische Astrophysik, Zentrum f\"{u}r Astronomie der Universit\"{a}t Heidelberg, Albert-Ueberle-Strasse 2, 69120 Heidelberg, Germany
    \and
    Cosmic Origins Of Life (COOL) Research DAO, coolresearch.io
    \and
    Department of Physics and Astronomy, University of Wyoming, Laramie, WY 82071, USA
    \and
    National Radio Astronomy Observatory, 520 Edgemont Road, Charlottesville, VA 22903, USA
    \and
    Research School of Astronomy and Astrophysics, Australian National University, Canberra, ACT 2611, Australia
    \and
    ARC Centre of Excellence for All Sky Astrophysics in 3 Dimensions (ASTRO 3D), Australia
    \and
    Visiting Fellow, Harvard-Smithsonian Center for Astrophysics, 60 Garden Street, Cambridge, MA 02138, USA
    \and
    Astrophysics Research Institute, Liverpool John Moores University, 146 Brownlow Hill, Liverpool L3 5RF, UK
    \and
    Max Planck Institute for Astronomy, Königstuhl 17, 69117 Heidelberg, Germany
    \and
    Centro de Desarrollos Tecnológicos, Observatorio de Yebes (IGN), 19141 Yebes, Guadalajara, Spain
    \and
    Sterrenkundig Observatorium, Universiteit Gent, Krijgslaan 281 S9, B-9000 Gent, Belgium
    \and
    Department of Physics and Astronomy, Rutgers University, 136 Frelinghuysen Road, Piscataway, NJ 08854, USA
    \and
    Department of Physics, Tamkang University, No.151, Yingzhuan Road, Tamsui District, New Taipei City 251301, Taiwan
    \and
    National Astronomical Observatory of Japan, 2-21-1 Osawa, Mitaka, Tokyo, 181-8588, Japan
    \and
    Center for Astrophysics and Space Sciences, University of California, San Diego, 9500 Gilman Drive MC0424, La Jolla, CA 92093, USA
    \and
    Sub-department of Astrophysics, Department of Physics, University of Oxford, Keble Road, Oxford OX1 3RH, UK
    }

   \date{Received February 5, 2024; accepted June 13, 2024}

\abstract{
The star formation rate (SFR) is tightly connected to the amount of dense gas in molecular clouds. 
However, it is not fully understood how the relationship between dense molecular gas and star formation varies within galaxies and in different morphological environments.
Most previous studies were typically limited to kpc-scale resolution such that different environments could not be resolved.
{In this work, we study dense gas and star formation in the nearby spiral galaxy NGC\,4321 to test how the amount of dense gas and its ability to form stars varies with environmental properties at 260\,pc scales.} 
We present new ALMA observations of HCN(1$-$0) line emission, which traces the dense gas content across the molecular gas disc at 260\,pc scales.
Combined with existing CO(2$-$1) observations from ALMA to trace the bulk molecular gas, and \halpha from MUSE, as well as F2100W from JWST to trace the SFR, we measure the HCN/CO line ratio, a proxy for the dense gas fraction and SFR/HCN, a proxy for the star formation efficiency of the dense gas.
Towards the centre of the galaxy, HCN/CO systematically increases while SFR/HCN decreases, but these ratios stay roughly constant throughout the disc. 
Spiral arms, interarm regions, and bar ends show similar HCN/CO and SFR/HCN.
On the bar, there is a significantly lower SFR/HCN at a similar HCN/CO.
We conclude that the centres of galaxies show the strongest environmental influence on dense gas and star formation (average variations of $>\SI{1}{\dex}$ in the SFR/HCN and HCN/CO ratios), suggesting either that clouds couple strongly to the surrounding pressure or that HCN is tracing more of the bulk molecular gas that is less efficiently converted into stars. 
On the contrary, across the disc of NGC\,4321, where the ISM pressure is typically low, SFR/HCN does not show large variations ($<\SI{0.3}{\dex}$) in agreement with Galactic observations of molecular clouds.
We suggest that gas dynamics, e.g. shear and strong streaming motions in galaxy bars, can have a large effect on the efficiency with which dense gas is converted into stars. 
Despite the large variations across environments and physical conditions, HCN/CO is a good predictor of the mean molecular gas surface density at 260\,pc scales.}


   \keywords{ISM: molecules --
             Galaxies: ISM --
             Galaxies: star formation --
             Galaxies: individual: NGC\,4321
               }

   \maketitle
%

\section{Introduction}

Galactic observations of dust in star-forming regions show that stars form in dense substructures, where the inferred star formation rate (SFR) is found to be linearly related to the amount of dense gas \citep[e.g.][]{Heiderman2010, Lada2010, Lada2012, Evans2014}.
\citet{Gao2004} found that this linear relation also holds for global measurements of galaxies when tracing the SFR with the total infrared (IR) luminosity and the dense gas mass (\mdense) via the luminosity of \hcnone.
Molecular line emission from HCN has an effective critical density of $\neff\sim\SI{5e3}{\per\cubic\cm}$, which is at least one order of magnitude higher than that of CO \citep[$\neff\lesssim\SI{e2}{\per\cubic\cm}$;][]{Shirley2015}.
Over the last two decades, many studies aimed at mapping HCN across other galaxies \citep[e.g.][]{Usero2015, Bigiel2016, Gallagher2018a, Jimenez-Donaire2019, Querejeta2019, Sanchez-Garcia2022, Neumann2023a}. These observations of star-forming, spiral galaxies and numerical works \citep[e.g.][]{Onus2018} tell us that the IR luminosity, tracing embedded SFR, is tightly (scatter of $\pm\SI{0.4}{\dex}$) and linearly correlated with the HCN luminosity, tracing \mdense, spanning ten orders of magnitude \citep[see e.g.][Be\v{s}li\'c et al. subm.; Schinnerer \& Leroy subm. for literature compilations]{Jimenez-Donaire2019, Neumann2023a}.

Despite the clear relation between SFR and dense gas, there is still a total scatter of $\approx\SI{1}{dex}$ that cannot solely be explained by measurement uncertainties, instead indicating that the dense gas star formation efficiency ($\text{SFR}/\mdense\equiv\sfedense$) depends on other physical quantities.
Over the last decade, resolved, kpc-scale observations of nearby galaxies \citep[e.g.][]{Usero2015, Bigiel2016, Gallagher2018a, Gallagher2018b, Jimenez-Donaire2019, Querejeta2019, Sanchez-Garcia2022, Neumann2023a}  have studied the variation of spectroscopic ratios, like HCN/CO, a proxy of the dense gas fraction ($\fdense\equiv \mdense/\mmol$, where \mmol is the molecular gas mass), and IR/HCN, a proxy of the dense gas star formation efficiency (\sfedense) with environmental properties, such as the stellar mass surface density (\sigstar), the molecular gas surface density (\sigmol) or the hydrostatic pressure in the ISM disc (\PDE).
These studies find that \fdense and \sfedense vary systematically with the environment. 
In particular, \fdense is significantly enhanced while \sfedense is systematically suppressed in high-surface density, high-pressure regions indicating a connection between the properties of molecular clouds and their host environment.
These results are also supported by studies of the Milky Way central molecular zone (CMZ), where \sfedense is found to be systematically lower than across the Milky Way disc \citep[e.g.][]{Longmore2013, Kruijssen2014b, Henshaw2023}.

In pioneering work, \citet{Gallagher2018b} find systematic correlations between the kpc-scale \fdense, \sfedense, and the molecular gas surface density measured at giant molecular cloud (GMC) scales, i.e. $\sim\SI{100}{\parsec}$.
Building upon this, \citet{Neumann2023a} use HCN observations of 25 nearby galaxies from the ``ACA Large-sample Mapping Of Nearby galaxies in Dense gas'' (ALMOND) survey in order to compare the kpc-scale spectroscopic line ratios with the properties of the molecular gas as traced by \cotwo on $\sim\SI{100}{\parsec}$ scales from the ``Physics at High ANgular resolution GalaxieS''--``Atacama Large Millimetre Array'' (PHANGS--ALMA) survey \citep{Leroy2021b}.
They showed that \fdense increases and \sfedense decreases with increasing surface density (\sigmol) and velocity dispersion (\vdis) of the molecular gas measured at GMC scales.
These results are also in agreement with predictions from models describing the star formation in turbulent clouds \citep[e.g.][]{Padoan2002, Krumholz2005, Krumholz2007} and the ISM disc structure \citep[e.g.][]{Ostriker2010}, hence yielding a coherent picture between dense gas, star formation and turbulent cloud models.
In particular, these results have shown that \sfedense is not universal but depends on the environment, and that density-sensitive line ratios like HCN/CO are powerful extragalactic tools to trace the underlying density structure at $\sim\SI{100}{\parsec}$ scale even if measured at kpc-scales.

Previous studies of the relationship between dense gas, star formation and environment \citep[e.g.][]{Usero2015, Gallagher2018a, Gallagher2018b, Jimenez-Donaire2019, Neumann2023a} were thus limited to mapping dense gas at kpc-scales.
There exist only a few $\sim\SI{100}{\parsec}$ resolution maps of HCN, or other dense gas tracers (\citealp{Kepley2014}, M82, \SI{200}{\parsec}; 
\citealp{Chen2017}, outer spiral arm of M51, \SI{150}{\parsec}; 
\citealp{Harada2018}, NGC\,3256, \SI{200}{\parsec}; 
\citealp{Viaene2018}, GMCs in M31; \SI{100}{\parsec}; 
\citealp{Kepley2018}, IC10, \SI{34}{\parsec}; 
\citealp{Querejeta2019}, M51, \SI{100}{\parsec}; 
\citealp{Harada2019}, circumnuclear ring of M83, \SI{60}{\parsec}; 
\citealp{Beslic2021}, NGC\,3627, \SI{100}{\parsec}; 
\citealp{Martin2021}, NGC\,253, \SI{250}{\parsec} with the potential of resolutions $<\SI{50}{\parsec}$;
\citealp{Eibensteiner2022}, central \SI{2}{\kilo\parsec} of NGC\,6946, \SI{150}{\parsec};
\citealp{Sanchez-Garcia2022}, NGC\,1068, \SI{60}{\parsec}; 
\citealp{Stuber2023}., M51, \SI{125}{\parsec}; 
Be\v{s}li\'c et al. subm., NGC\,253, \SI{300}{\parsec}).
However, these are typically less sensitive and they target certain regions but not the full disc, compared to the observations presented here.
Many of these works that mapped the whole molecular gas disc did not detect much emission in individual sight lines outside of galaxy centres and hence had to average over larger regions, e.g. via spectral stacking \citep[e.g.][]{Schruba2011, Caldu-Primo2013, Jimenez-Donaire2017, Jimenez-Donaire2019, Neumann2023b}, at the cost of spatial information.
Apart from a few exceptions (M51, NGC\,253, see references above), there are no deep, wide-field studies of these dense gas ratios at sub-kpc scales, which detect individual sight lines in different morphological environments across the whole molecular gas disc out to \SI{10}{\kilo\parsec} in galactocentric radius.
Such a study is, however, needed to study the sub-kpc structured ISM without blending many regions together that may have substantially different environmental and dynamical conditions for the formation of dense gas and its conversion to stars.
We need $\lesssim\SI{500}{\parsec}$ scale resolution in order to resolve environments like the centre (size of $\sim\SI{1}{\kilo\parsec}$), spiral arms (width of $\sim\SI{1}{\kilo\parsec}$) and bar ends (size of $\sim 0.5-\SI{1}{\kilo\parsec}$).

In this work, we present new ALMA observations of dense molecular gas tracers, including \hcnone, \hcopone, and \cstwo, across the nearby spiral galaxy \gal at $\ang{;;3.5}\sim\SI{260}{\parsec}$ resolution covering the full disc, i.e. out to $1.1\, r_{25}$.
These data are paired with \cotwo observations from PHANGS--ALMA \citep{Leroy2021b}, tracing the bulk molecular gas, and extinction corrected \halpha from PHANGS--MUSE \citep{Emsellem2022}, as well as \SI{21}{\micron} observations from PHANGS--JWST \citep{Lee2023} and \SI{33}{\giga\hertz} observations from VLA \citep{Linden2020}, used to trace the SFR.
These give us one of the best high-resolution, high-sensitivity data sets combining interferometric and total power observations of high critical-density molecular line emission accompanied by the most robust tracers of SFR across the full disc of a nearby star-forming galaxy.

The paper is structured as follows. In Sect.~\ref{sec:data}, we present the observations and ancillary data of \gal used in this work, including new ALMA HCN observations. In Sect.~\ref{sec:methods}, we describe the methods to derive physical quantities from the observations, including the dense gas content, star formation rate and ISM pressure. Then, in Sect.~\ref{sec:results}, we show the results, where we analyse the dense gas spectroscopic line ratios and their variation with environment, which are discussed in Sect.~\ref{sec:discussion}.
Finally, we conclude and summarise the key findings in Sect.~\ref{sec:conclusions}.

\section{Observations}
\label{sec:data}

\begin{figure*}
\centering
\includegraphics[width=\textwidth]{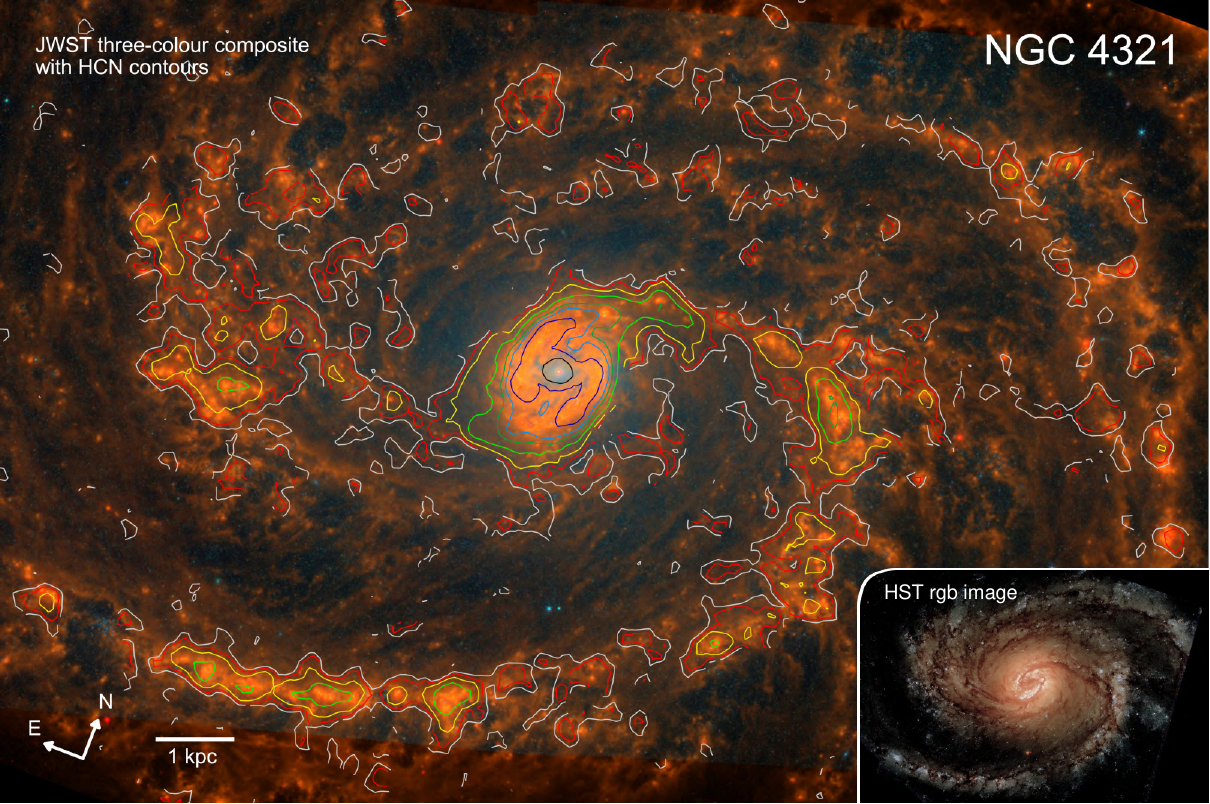}
\caption{JWST three-colour image of \gal overlaid with HCN contours. The background image is a three-colour composite using the MIRI and NIRCAM instruments observations (red=F770W+F1000W+F1130W+F2100W, green=F360M+F770W, and blue=F300M+F335M) taken from the PHANGS--JWST treasury survey \citep{Lee2023}.
Overlaid \hcnone contours (new data presented in this work), tracing the dense molecular gas are drawn at S/N levels of $(2, 3, 5, 10, 20, 30, 50, 100)$. The sites of star formation (reddish hues) appear spatially well correlated with the dense gas traced by HCN. This image is rotated by 21$^{\circ}$ with respect to the right ascension-declination plane as indicated by the north (N)-east (E) coordinate axes in the bottom left.
The bottom right image shows a rgb image (red=F814W, green=F555W, blue=F438W+F336W+F275W) from PHANGS--HST \citep{Lee2022}.}
\label{fig:jwst_map}
\end{figure*}


\begin{figure*}
\centering
\includegraphics[width=\textwidth]{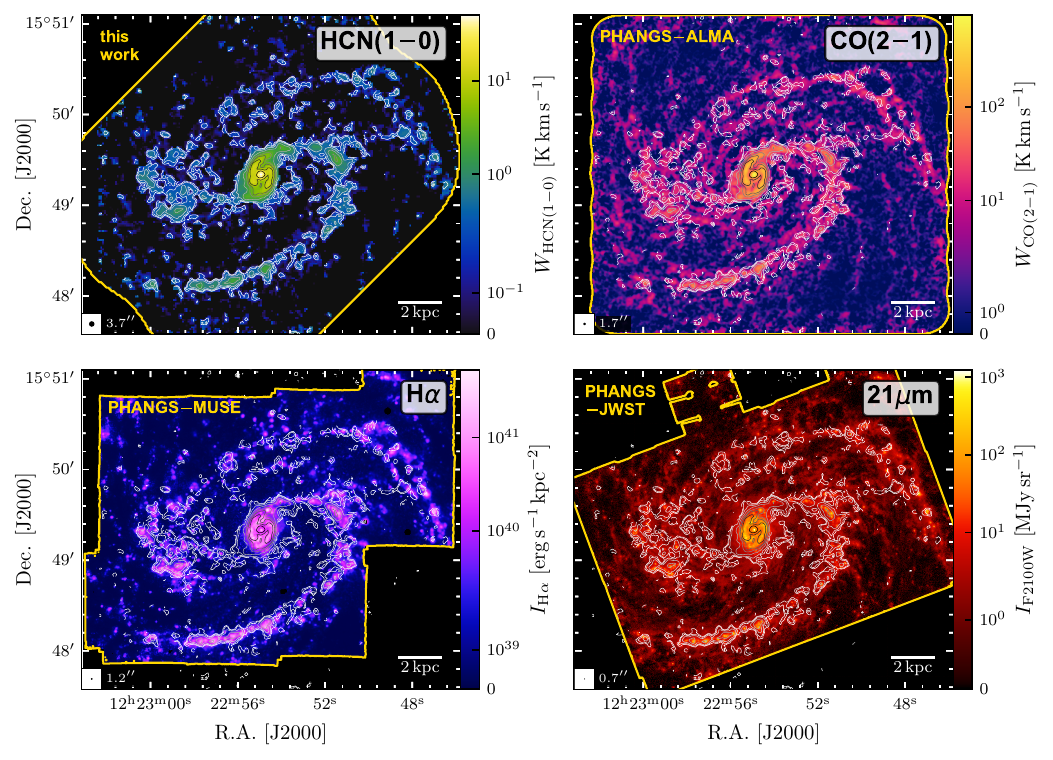}
\caption{\gal data used in this study, each at the native resolution of the respective observations indicated in the bottom left of each panel. \textit{Top left:} \hcnone moment-0 map presented in this work. \textit{Top right:} \cotwo moment-0 map from PHANGS--ALMA \citep{Leroy2021b}. \textit{Bottom left:} Extinction-corrected \halpha flux density from PHANGS--MUSE \citep{Emsellem2022}. \textit{Bottom right:} \SI{21}{\micron} flux density from MIRI-F2100W (PHANGS--JWST; \citealp{Lee2023}). In each panel, white-to-black-gradient contours show HCN moment-0 signal-to-noise ratio levels of $(2, 3, 5, 10, 20, 30, 50, 100)$ as in Fig.~\ref{fig:jwst_map}. The yellow-coloured outline shows the FOV of the respective observations.}
\label{fig:maps_observations}
\end{figure*}

\subsection{The target -- NGC\,4321}
We selected \gal for this study as previous ALMA/IRAM mapping showed clear HCN detections, supporting data covers almost all aspects of the ISM and galactic structure, and its favourable distance to obtain a wide area map while still resolving, e.g., the galactic centre, bar, arms and other regions.
\gal (main properties listed in Table~\ref{tab:target}) is a well-studied, spiral, barred \citep{Querejeta2021} galaxy (Hubble classification: SABbc) that contains a large reservoir of molecular gas ($M_{\htwo}=\SI{7.77e9}{\Msun}$; \citealp{Leroy2021b}), is actively forming stars ($\mathrm{SFR}=\SI{3.56}{\Msun\per\year}$; \citealp{Leroy2019}) and can be observed relatively face-on ($i=\SI{38.5}{\degree}$; \citealp{Lang2020}).
Fig.~\ref{fig:jwst_map} shows a JWST three-colour image overlaid with HCN contours from this work that highlight the spiral arm structure of the galaxy, seen in dust, gas and star formation. 
At a distance of $d=\SI{15.2}{\mega\parsec}$ \citep{Anand2021} it is relatively nearby, allowing access to GMC scales ($<\SI{100}{\parsec}$) at $\sim\ang{;;1}$ angular resolution. 
Moreover, \gal is a spiral galaxy with similar stellar mass ($M_\star=\SI{5.6e10}{\Msun}$) to our Galaxy ($M_\star=\SI{6.1e10}{\Msun}$, \citealp{Licquia2015}), making it an interesting object to compare with Galactic studies.
\gal has been extensively studied as part of large observing campaigns like PHANGS--ALMA \citep{Leroy2021b}, mapping \cotwo across the full disc of the galaxy at $\sim\ang{;;1}\sim\SI{100}{\parsec}$ resolution, as well as EMPIRE \citep{Jimenez-Donaire2019} and ALMOND \citep{Neumann2023a} mapping various dense gas tracers including HCN and HCO$^+$ with the IRAM 30-m telescope and the Atacama Compact Array (ACA), respectively, at kpc scales. 
Furthermore, \gal was part of the ALMA science verification \coone observations \citep{Pan2017} and has high-quality maps of \hone (HERACLES;  \citealp{Leroy2009}), stellar structure \citep[S$^4$G;][]{Sheth2010}, star formation tracers (\halpha from MUSE; \citealp{Emsellem2022}), as well as near and mid-infrared maps from the James Webb Space Telescope \citep[JWST;][]{Lee2023}.
We show a compilation of the key observations used in this work in Figure~\ref{fig:maps_observations}.

\begin{table}
    \begin{center}
    \caption{Properties of NGC\,4321.}
    \label{tab:target}
    \begin{tabular}{l l}
    \hline \hline
     Property    & Value  \\ \hline
     Alternative Name         & M100  \\
     Right Ascension (J2000)$^{\rm (a)}$ & \ra{12;21;54.9} \\
     Declination (J2000)$^{\rm (a)}$ & \ang{+04;28;25.5} \\
     Inclination, $i^{\rm (b)}$ & $(38.5\pm 2.4)^\circ$ \\
     Position Angle$^{\rm (b)}$ & $(156.2\pm 1.7)^\circ$ \\
     Radius, $r_{25}^{\rm (d)}$ & $(182.9\pm 47.3)$\arcsec \\
     Systemic Velocity, $V_{\rm LSR}^{\rm (b)}$ & $(1572\pm 5)\,$km\,s$^{-1}$\\
     Distance, $d^{\rm (a)}$ & $(15.21\pm 0.49)$\,Mpc \\
     Linear Scale & 73.5\,pc/\arcsec \\
     Matched Beam Size & $\SI{3.5}{\arcsec} \sim \SI{260}{\parsec}$ \\
     Morphology$^{\rm (e)}$ & SAB(s)bc \\
     ${\rm SFR}^{\rm (c)}$  & $(3.56\pm 0.92)$\, M$_\odot$\,yr$^{-1}$\\
     $\log_{10}(M_\star/\mathrm{M}_\odot)^{\rm (c)}$ & $10.75\pm 0.11$ \\ \hline
    \end{tabular}
     \end{center}
    \raggedright{ {\bf Notes:}\\
    (a) \cite{Anand2021};\\ 
    (b) \cite{Lang2020};\\
    (c) \cite{Leroy2019};\\
    (d) HyperLeda database \citep{Makarov2014};\\
    (e) NASA Extragalactic Database (NED).}
\end{table}

\subsection{New ALMA maps of HCN}
\label{sec:data:hcn}
In this work, we present ALMA Band-3 observations (2017.1.00815.S; PI.: Molly Gallagher) which mapped \hcnone (along with \hcopone and \cstwo) across the full disc of the galaxy \gal at a high angular resolution of \ang{;;3.5} using \SI{216.7}{\hour} of ALMA telescope time.
The observations combine interferometric observations from the 12-m array (\SI{18.1}{\hour} observing time) with the ACA  consisting of the 7-m array (\SI{73.4}{\hour}) and the 12-m dishes observing in total power (TP) mode (\SI{125.2}{\hour}).
The mapped area on the sky is $\ang{;;200}\times\ang{;;120}$ large created via a mosaic consisting of 27 Nyquist-spaced pointings with the 12-m array.
The spectral setup encompasses four spectral windows, each with a bandwidth of \SI{1875}{\mega\hertz} and a channel width of $\SI{976}{\kilo\hertz}$.
The first window, centred at \SI{88.5}{\giga\hertz} targets \hcnone\, (88.6\,GHz) and \hcopone\, (89.2\,GHz).
The second window at \SI{87.0}{\giga\hertz} covers SiO(2--1) (86.9\,GHz) and isotopologues of HCN and HCO$^+$, i.e. H$^{13}$CN\,(1--0), H$^{13}$CO$^+$\,(1--0).
The third spectral window at \SI{98.5}{\giga\hertz} comprises \cstwo (97.9809533\,GHz).
The fourth spectral window at \SI{100}{\giga\hertz} is used to detect continuum emission.
The channel width of $\approx\SI{3}{\kilo\metre\per\second}$ is sufficient to resolve the spectral lines across the whole disc of the galaxy, and the bandwidth of $\approx\SI{6000}{\kilo\metre\per\second}$ allows mapping all lines over the full velocity extent. 

The data reduction was performed using the PHANGS--ALMA pipeline (details can be found in \citealp{Leroy2021a}) which utilises the standard ALMA data reduction package \texttt{CASA} \citep{casa_package}. 
In this first study, we focus on \hcnone (hereafter HCN) as the brightest proxy for dense molecular gas.
The resulting HCN position-position-velocity cube has $\sim\SI{8}{\milli\kelvin}$ noise per \SI{5}{\kilo\metre\per\second} channel.
The high resolution, which corresponds to \SI{260}{pc} physical scales allows us to resolve individual environmental regions including the centre, bar, bar ends, spiral arms and interarm regions (Fig.~\ref{fig:hcn_ratio_maps}, right panel) yielding detection of 302 independent lines of sight in HCN emission (see Sect.~\ref{sec:methods:mom0_maps} for details on masking and derivation of moment-0 maps).

\subsection{Ancillary data}
\label{sec:data:ancillary_data}
In addition to the new HCN data, tracing dense molecular gas, we use CO observations to trace the bulk molecular gas (Sect.~\ref{sec:methods:sigmol}) and \halpha observations to trace SFR (Sect.~\ref{sec:methods:sfr_halpha}).
Furthermore, we include \hone 21-cm observations (Sect.~\ref{sec:appendix:hi}) and \SI{3.6}{\micron} infrared maps (Sect.~\ref{sec:appendix:mstar}) to trace the atomic gas and the stellar mass content, respectively.
In the Appendix, we further present additional tracers of the SFR, i.e. F2100W hot dust observations from JWST (Sect.~\ref{sec:appendix:sfr_21mu}) and \SI{33}{\giga\hertz} free-free emission from the VLA (Sect.~\ref{sec:appendix:sfr_33ghz}), supporting the use of \halpha as the primary SFR tracer in this work.

\section{Methods}
\label{sec:methods}

\subsection{Integrated intensity maps}
\label{sec:methods:mom0_maps}
We produce integrated intensity maps (moment-0 maps) from the CO and HCN position-position-velocity (PPV) cubes following \citet{Neumann2023a}. 
The methodology goes back to \citet{Schruba2011} and was utilised in several studies such as EMPIRE \citep{Jimenez-Donaire2019}, CLAWS \citep{denBrok2022} and ALMOND \citep{Neumann2023a}.
First, we homogenise the data by convolving the CO data to the HCN resolution (using \texttt{convolution.convolve} from \texttt{astropy}).
Then, we adopt a hexagonal spaxel grid with a beam-size spaxel separation and sample all data to the same spaxel grid and spectral axis.
This means that every hexagonal pixel is an independent line-of-sight (LOS) measurement.
Then, we create velocity masks based on the CO on a pixel-by-pixel basis to select the velocity range where we also expect to find HCN emission.
This is done by building a $4\sigma$ mask that is expanded into channels above $2\sigma$ in order to recover broader emission belonging to a $4\sigma$ core \citep[see e.g.,][for more details about the masking]{Neumann2023a}.
By applying the CO-based mask to our data, we compute the integrated intensity of CO ($W_{\rm CO}$) and HCN ($W_{\rm HCN}$) by integrating the line's brightness temperatures ($T_{\rm line}$, where ${\rm line}=\{\mathrm{CO}, \mathrm{HCN}\}$) over the velocity range selected by the mask:
\begin{align}
    \left(\dfrac{W_{\rm line}}{\SI{}{\Kkms}}\right) = \sum_{n=1}^{N_{\rm mask}} \left(\dfrac{T_{\rm line, n}}{\SI{}{\kelvin}}\right) \times \left(\dfrac{\Delta v_\text{channel}}{\SI{}{\kilo\metre\per\second}}\right) \;.
\end{align}
The uncertainties of the integrated intensities ($\sigma_{W_{\rm line}}$) are then given by:
\begin{align}
    \left(\dfrac{\sigma_{W_{\rm line}}}{\SI{}{\Kkms}}\right) = \left(\dfrac{\sigma_{T_{\rm line}}}{\SI{}{\kelvin}}\right) \times \left(\dfrac{\Delta v_\mathrm{channel}}{\SI{}{\kilo\metre\per\second}}\right) \times \sqrt{N_{\rm mask}} \;,
    \label{equ:emom0}
\end{align}
where $\sigma_{T_{\rm line}}$ is the standard deviation in the emission-free channels (i.e. channels not selected by the mask), $\Delta v_\mathrm{channel}$ is the channel width of \SI{5}{\kilo\metre\per\second} and $N_{\rm mask}$ is the number of channels selected by the mask for each LOS.

We note that we also homogenise the two-dimensional maps, e.g. the MUSE \halpha and JWST \SI{21}{\micron} maps, with the produced moment-0 maps.
This means, we convolve the maps to the \SI{260}{\parsec} HCN resolution and reproject them onto the same beam-size hexagonal pixel grid.
A summary of the data products is presented in Fig.~\ref{fig:data_products_maps}.
We describe the derivation of the physical quantities in the following subsections.

\subsection{Molecular gas surface density -- CO}
\label{sec:methods:sigmol}
We use \cotwo (hereafter CO) line observations from PHANGS--ALMA \citep{Leroy2021b} to trace the bulk molecular gas.
For \gal, the CO data are at \ang{;;1.67} resolution, which corresponds to \SI{120}{\parsec} physical scale at the distance of the galaxy.
\footnote{We do not use the archival \coone observations as a tracer of molecular gas because of their poorer angular resolution (\ang{;;4}, corresponding to $\sim 300$~pc), but we do use them to inform our conversion of \cotwo intensity to molecular gas surface density, as described later in Sect.~\ref{sec:methods:sigmol}.}
We infer the molecular gas surface density (\sigmol) from the \cotwo line intensity (\intCO) using the \cotwo-to-\coone line ratio ($R_{21}$) and the CO-to-\htwo conversion factor (\aCO), which includes the mass contribution from helium:
\begin{equation}
    \sigmol = \aCO \, R_{21}^{-1} \, \intCO \cos(i).
\end{equation}
$\cos(i)$ corrects for the inclination $i=\ang{38.5}$ of the galaxy.
Throughout this work, we adopt two methods (see Appendix~\ref{sec:appendix:aco} for more details): 1) using constant \aCO and $R_{21}$ conversion factors (Sect.~\ref{sec:methods:sigmol_const_aco}) which enter the estimation of the dense gas fraction as traced by the HCN-to-CO line ratio (Sect.~\ref{sec:methods:fdense}). 
We use a constant \aCO for the HCN-to-CO line ratio due to the poor knowledge about variations of the HCN-to-dense gas conversion factor thus keeping \fdense proportional to HCN/CO. 
2) using spatially varying \aCO and $R_{21}$ (Sect.~\ref{sec:methods:sigmol_const_aco}) for computing \sigmol and the dynamical equilibrium pressure (Sect~\ref{sec:methods:pde}).

\subsection{Dense gas fraction -- HCN/CO}
\label{sec:methods:fdense}
In this study, we present new HCN observations (Sect.~\ref{sec:data:hcn}) and use the HCN line intensity (\intHCN) as a proxy for the amount of dense gas.
For the main part of this work, we focus on studying the observational HCN-to-CO line ratio, i.e. \intHCN/\intCO (hereafter HCN/\cotwo or simply HCN/CO) as a density-sensitive line ratio.
\citet{Gallagher2018b} and \citet{Neumann2023a} have shown that HCN/CO is indeed tracing the $\sim\SI{100}{\parsec}$-scale mean gas density and it has been reported to scale with the gas surface density within Galactic clouds \citep{Tafalla2023} as expected by molecular line modelling \citep{Leroy2017a}.
In addition, the reported linear relation between the HCN/CO and \nnhp/CO line ratios across Galactic and extragalactic studies underlines the credibility of HCN/CO as a proxy for the dense gas fraction \citep{Jimenez-Donaire2023, Stuber2023}.
Throughout the discussion (Sect.~\ref{sec:discussion}), we comment on the implications of the dense gas fraction (\fdense) as a physical quantity proportional to HCN/CO with some uncertainties linked to abundance, temperature and opacity.

Following many previous works \citep[e.g.][]{Usero2015, Bigiel2016, Gallagher2018a, Gallagher2018b, Jimenez-Donaire2019, Bemis2019, Neumann2023a}, the dense gas fraction is defined as the ratio of the dense gas to bulk molecular gas surface density ($\fdense=\sigdense/\sigmol$):
\begin{align}
    \fdense = \dfrac{\sigdense}{\sigmol} = \dfrac{\aHCN\intHCN}{\aCO R_\text{21}^{-1}\intCO} \approx 2.1\dfrac{\intHCN}{\intCO}  \, .
\end{align}
The above conversion adopts constant mass-to-light ratios $\aCO=\SI{4.35}{\Msun\per\square\parsec}\,(\SI{}{\Kkms})^{-1}$ \citep{Bolatto2013} and $\aHCN=\SI{14}{\Msun\per\square\parsec}\,(\SI{}{\Kkms})^{-1}$ \citep{Onus2018} for CO and HCN, respectively, and a fiducial \cotwo-to-\coone line ratio of $R_{21}=0.65$ \citep{denBrok2022, Leroy2022}.
Here, we use the above conversion to infer \fdense as an alternative axis in the HCN/CO relations. 

The adopted constant HCN-to-dense gas mass conversion factor is expected to trace gas above $n_{{\rm H}_2}\approx\SI{5e3}{\per\cubic\centi\metre}$ \citep{Onus2018}\footnote{Note that many previous works \citep[e.g.][]{Gao2004} use a slightly smaller value of $\aHCN=\SI{10}{\Msun\per\square\parsec}\,(\SI{}{\Kkms})^{-1}$ tracing gas above $n_{{\rm H}_2}\approx\SI{3e4}{\per\cubic\centi\metre}$.
However, choosing a different (constant) \aHCN has no qualitative effect on our results.}.
In contrast to \aCO, systematic variations of \aHCN are poorly understood and estimated values range from 0.3 to $\SI{300}{\Msun\per\square\parsec}\,(\SI{}{\Kkms})^{-1}$, spanning three orders of magnitude \citep{Garcia-Burillo2012, Kauffmann2017, Nguyen-Luong2017, Shimajiri2017, Evans2020, Barnes2020, Tafalla2023}, where extragalactic studies, capturing larger physical areas and thus more diffuse emission typically yield values around $\SIrange{10}{20}{\Msun\per\square\parsec}\,(\SI{}{\Kkms})^{-1}$.
The \aHCN conversion factor might vary similarly to \aCO due to its dependence on optical depth, which is a key driver of \aCO variations \citep{Teng2023}, though HCN and CO optical depth variations are not expected to be identical.
In that case, we could even induce systematic trends by adopting a more accurate, spatially varying \aCO, but keeping \aHCN constant.
Therefore, the best current approach is to study the observational HCN/CO line ratio.

As laid out in this section, we adopt the classical view of utilising HCN/CO as a proxy of \fdense.
However, we want to point out that the conversion factors are subject to large uncertainties (especially \aHCN) such that our \fdense estimates are expected to be uncertain by a factor of a few.
Therefore, recent works \citep[e.g.][]{Gallagher2018a, Gallagher2018b, Jimenez-Donaire2019, Neumann2023a, Tafalla2023}, which study HCN/CO as a function of the molecular gas surface density suggest to interpret HCN/CO as a predictor of the cloud-scale ($\sim\SI{100}{\parsec}$) average gas density based on the robust relation between HCN/CO and \sigmol (see also Sect.~\ref{sec:discussion:hcn_co}).
This means, HCN/CO is expected to track \sigmolavg more robustly than \fdense.
We note, that in turbulent cloud models \citep{Krumholz2005}, an increase of HCN/CO would indicate an increase in \fdense as well as \sigmolavg.
Therefore, both interpretation, i.e. HCN/CO traces \fdense and HCN/CO traces \sigmolavg, are reasonable.
Throughout this work, we base our results on the observable HCN/CO line ratio, and provide a secondary \fdense-axis in Figures~\ref{fig:hcn_ratios_vs_radius_and_pressure} to \ref{equ:hcn_co_vs_sigmol}, so that, taking into considering aforementioned caveats, HCN/CO can be interpreted as the dense gas fraction via a proportional conversion or alternatively as an indicator of the mean gas density.

\subsection{Star formation rate (SFR) -- \halpha}
\label{sec:methods:sfr_halpha}
We use \halpha recombination line emission taken by the Multi Unit Spectroscopic Explorer (MUSE) of the Very Large Telescope (VLT) as part of the PHANGS--MUSE survey \citep{Emsellem2022} to trace the star formation rate.
In Appendix~\ref{sec:appendix:sfr_tracers}, we discuss using alternative SFR tracers including \SI{21}{\micron} (F2100W) hot dust emission from JWST \citep{Lee2023} and \SI{33}{\giga\hertz} free-free emission from the VLA \citep{Linden2020}, which can differ significantly (up to one order of magnitude) in the central few kiloparsecs of galaxies.
Here, we find that SFR values inferred from the \SI{33}{\giga\hertz} emission confirm the extinction-corrected \halpha inferred values in the centre of \gal. 
Moreover, \SI{21}{\micron} emission also yields similar SFR values (within \SI{0.2}{\dex}) when adopting a linear conversion (for more details see App.~\ref{sec:appendix:sfr_comparison}).
Therefore, throughout this work we adopt \halpha emission as a robust tracer of SFR validated by free-free data in the centre.

We use Balmer decrement-corrected \halpha maps to measure the star formation rate surface density (\sigsfr). 
Those rely on the extinction curve from \citet{O'Donnell1994} as described in \citet{Pessa2022} and \citet{Belfiore2023}. 
The attenuation corrected \halpha flux ($L_{\halpha,\mathrm{corr}}$) is converted into SFR via $\mathrm{SFR}/(\SI{}{\Msun\per\year})=C_{\halpha}\times L_{\halpha,\mathrm{corr}}/(\SI{}{\erg\per\second})$ using the conversion factor $C_{\halpha}=\num{5.3e-42}$ from \citet{Calzetti2007}.
This conversion assumes a constant star formation history, age of \SI{100}{\mega\year}, solar metallicity, and a \citet{Kroupa2001} initial mass function (IMF) \citep[for more detail on the SFR calibration, see][]{Belfiore2023}.
In surface density units the above formalism translates to:
\begin{equation}
    \left(\dfrac{\Sigma_{\rm SFR, \halpha}}{\SI{}{\Msun\per\year\per\square\kilo\parsec}}\right) = \num{6.3e2} \left( \dfrac{I_{\halpha, \rm corr}}{\SI{}{\erg\per\second\per\square\centi\metre\per\steradian}} \right) \cos(i) \;.
    \label{equ:sfr_halpha}
\end{equation} 

\subsection{Dense gas star formation efficiency -- SFR/HCN}
\label{sec:methods:sfedense}
We take the ratio of the star formation rate surface density (\sigsfr) to the HCN line intensity (\intHCN), i.e. \sigsfr/\intHCN (hereafter SFR/HCN) as a proxy of the star formation efficiency of the dense gas (\sfedense).
Similar to HCN/CO tracing \fdense, we also focus on the more observationally based SFR/HCN in our analysis and discuss implications on the inferred \sfedense connected to uncertainties in the conversion factor (\aHCN).
\sfedense is defined as the ratio of SFR surface density to dense gas mass surface density ($\sfedense=\sigsfr/\sigdense$) as in previous works (listed in Sect.~\ref{sec:methods:fdense}):
\begin{align}
    \sfedense = \dfrac{\sigsfr}{\sigdense} = \aHCN^{-1}\,\dfrac{\sigsfr}{\intHCN} \, .
    \label{EQU:sfedense}
\end{align}
The above equation yields 
\begin{align}
    \left(\dfrac{\sfedense}{\SI{}{\per\mega\year}}\right) = 0.071 \left(\dfrac{\sigsfr}{\SI{}{\Msun\per\year\per\square\kilo\parsec}}\right) \left(\dfrac{\intHCN}{\SI{}{\Kkms}}\right)^{-1} \, .
\end{align}
when using the same, constant HCN-to-dense gas mass conversion ($\aHCN=\SI{14}{\Msun\per\square\parsec}(\SI{}{\Kkms})^{-1}$, \citealp{Onus2018}) as for \fdense (Sect.~\ref{sec:methods:fdense}).

\begin{figure*}
\centering
\includegraphics[width=\textwidth]{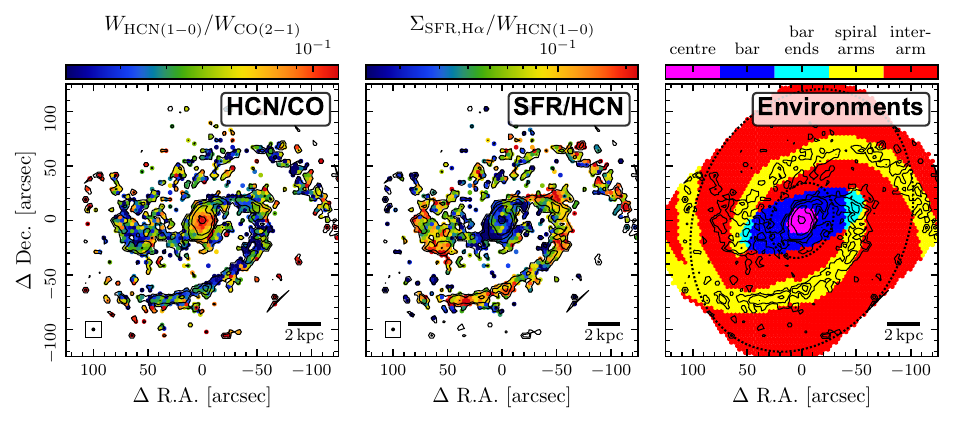}
\caption{\gal maps sampled at beam size. \textit{Left:} HCN/CO line ratio map, as a proxy of the dense gas fraction (\fdense). The \cotwo data are taken from the PHANGS-ALMA survey \citep{Leroy2021b}. \textit{Middle:} SFR surface density-to-HCN line intensity ratio, as a proxy of the dense gas star formation efficiency (\sfedense). The SFR surface densities are obtained from the Balmer decrement-corrected \halpha flux computed from PHANGS-MUSE data \citep{Emsellem2022, Belfiore2023}. Black contours show HCN S/N levels of 3, 10, 30, 100, and 300, in each of the panels. In the left and middle plot, only pixels above 1-sigma are shown. The colour scale goes from the 10th to the 95th percentile of pixels above 1-sigma noise level. \textit{Right:} Environmental region mask from \citet{Querejeta2021}, which is created based on the Spitzer \SI{3.6}{\micron} maps tracing the stellar mass content. Here, we select the regions centre, bar, bar ends, spiral arms and interarm, where the interarm includes the interbar region. The dotted ellipses show loci of constant galactocentric radius (\rgal), drawn at $\rgal=\SI{2.6}{\kilo\parsec}$ and $\rgal=\SI{9.17}{\kilo\parsec}$, where the latter indicates the largest radius completely covered by the HCN field of view.}
\label{fig:hcn_ratio_maps}
\end{figure*}

\subsection{Dynamical equilibrium pressure}
\label{sec:methods:pde}
We compute the dynamical equilibrium pressure, or ISM pressure (\PDE) at \SI{260}{\parsec} scale following the prescription by \citet{Sun2020a}.
In this prescription the dynamical equilibrium pressure is composed of a pressure term created by the ISM due to the self-gravity of the ISM disc and a term due to the gravity of the stars (see e.g. \citealp{Spitzer1942}), such that:
\begin{align}
    \PDE = \dfrac{\pi G}{2} \siggas^2 + \siggas \sqrt{2G\rhostar}\, \sigma_{\text{gas,z}} \; ,
    \label{equ:PDE_large}
\end{align}
where we assume a smooth, single-fluid gas disc, and that all gas shares a similar velocity dispersion, so that $\siggas=\sigmol +\sigatom$ is the total gas surface density, composing of a molecular (\sigmol) and an atomic (\sigatom)
gas component.
\rhostar is the stellar mass volume density (App.~\ref{sec:appendix:mstar}) near the disc mid-plane and $\sigma_{\text{gas,z}}$ is the velocity dispersion of the gas perpendicular to the disc.

In many previous extragalactic studies \citep[e.g.][]{Spitzer1942, Elmegreen1989, Elmegreen1994, Wong2002, Blitz2004, Blitz2006, Leroy2008, Koyama2009, Ostriker2010, Ostriker2011, Kim2011, Shetty2012, Kim2013, Kim2015, Benincasa2016, Herrera-Camu2017, Gallagher2018b, Fisher2019, Schruba2019, Jimenez-Donaire2019} \PDE was typically estimated using Equ.~\eqref{equ:PDE_large} with homogenised \siggas, \rhostar, $\sigma_{\text{gas,z}}$ at kpc scales. 
Recently, \citealp{Sun2020a} came up with a new formalism which makes use of the high resolution $\sim\SI{100}{\parsec}$ scale \cotwo data from PHANGS--ALMA.
Most importantly, it takes into account the self-gravity of the (molecular) gas at high resolution. 
In this study we adopt their formalism and combine the \SI{120}{\parsec} scale molecular gas term (\Pcloudavg; converted to the lower resolution via a \sigmol-weighted average) with the \SI{260}{\parsec} scale atomic gas term (\Patom):
\begin{align}
    \PDEavg = \Pcloudavg + \Patom.
    \label{equ:PDE_avg}
\end{align}
\Pcloudavg consists of three terms accounting for the self-gravity of the molecular gas, the gravity of larger molecular structures and the gravity of stars.
\Patom includes the self-gravity of the atomic gas and the gravitational interaction of the atomic gas with the \SI{260}{\parsec} scale molecular gas and the stars (see App.~\ref{sec:appendix:pde} for more details).

\subsection{Morphological environmental masks}
\label{sec:methods:environments}
We adopt the environmental masks presented in \citet{Querejeta2021}, which identify morphological environmental regions based on the appearance of the stellar mass content traced by the \textit{Spitzer} \SI{3.6}{\micro\metre} emission from S$^4$G \citep{Sheth2010}.
We use the ``simple'' mask, where each pixel is uniquely assigned to a dominant environment. 
We define the bar ends as the overlap of the spiral arms with the bar footprint.
For simplicity, we combine interbar, interarm into one region, referred to as interarm.
We end up with five environments -- centre, bar, bar ends, spiral arms, interam, which are re-sampled onto the same hexagonal grid as the other data defined by the HCN map (Sect.~\ref{sec:methods:mom0_maps}). 
We show the adopted environments in the right panel of Fig.~\ref{fig:hcn_ratio_maps}.

\subsection{Stacking and linear regression}
\label{sec:methods:fitting}
\begin{figure*}
\centering
\includegraphics[width=\textwidth]{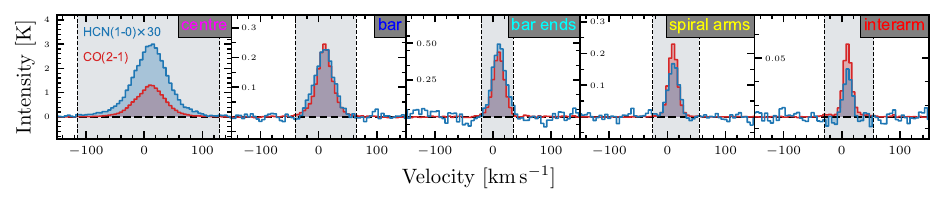}
\caption{Stacked spectra by environment. The \cotwo line (red) is used to correct for the effect of the velocity field. In every panel, the \hcnone intensities are multiplied by a factor of 30 for better comparison with the CO intensities. The grey-shaded area indicates the velocity window over which the integrated intensities are computed.}
\label{fig:stacks_envir}
\end{figure*}

To study the average trends, we stack the data (HCN, CO, SFR) in equally spaced bins in linear scale (\rgal) or logarithmic scale (\PDE).
The spectral stacking is done using the \texttt{python} package \texttt{PyStacker}\footnote{\url{https://github.com/PhangsTeam/PyStacker}}, which yields average CO and HCN spectra in each respective bin (the stacked spectra are presented in Fig.~\ref{fig:stacks_envir} and in App.~\ref{sec:appendix:stacking}).
Those average spectra are used to compute the average integrated intensities in each bin.
For SFR, we simply compute the mean of the SFR values in the respective bin.
The bin ratios are then computed as the ratios of the stacked measurements.
To first order, HCN and CO lines show similar kinematics across most of the galaxy, so the line ratio, which we discuss in this first paper, encodes most of the relevant information.

We fit the stacked data in order to probe the underlying global relation without ``population'' biases and to not be dominated by non-detections in constraining the best-fit line.
We note, however, that we have also fitted the individual sight-line measurements using \texttt{LinMix}\footnote{\url{https://linmix.readthedocs.io/en/latest/index.html}; \texttt{LinMix} is a Bayesian inference tool to linear regression, which can take into account upper limits and infers the posterior distribution of the fit line parameters via MCMC simulation \citep{Kelly2007}.} resulting in similar fit relations in agreement within 1-sigma uncertainties with the fits reported here for the binned data.
We use these sight-line fits to quantify the uncertainty of the regression slopes since the MARS fitting routine (see below) does not yield uncertainties.

We then apply a `multivariate adaptive regression spline' (MARS; \citealp{Friedman1991}) model to the binned data in order to find the best piece-wise linear regression function that describes the data (see Sect.~\ref{sec:results:radius} and \ref{sec:results:pressure}).
MARS is a generalisation of a recursive partitioning algorithm, which iteratively splits the data into separate $x$-axis regimes and optimises the split point with respect to the the piece-wise linear regression in each regime via minimising the $\chi^2$ value of the data to the model.
The algorithm is adapted to only add another split point if a further component significantly improves the fit, meaning that the $\chi^2$ value is improved by more than 0.01.
In this way, we employ a statistically robust and objective method to find the threshold at which the trends change significantly thus identifying physically different regimes in the relations.
To perform the MARS model we utilise the \texttt{R}-package \texttt{earth}\footnote{\url{https://cran.r-project.org/package=earth}}.
Here, we force the model to only consist of up to two linear functions, i.e. it can either find one or two regimes depending on if a second regime improves the fit significantly.

\begin{table*}
    \begin{center}
    \caption{Line intensities and luminosities by environment.}
    \label{tab:stacking_values}
    \begin{tabular}{ccccccc}
    \hline \hline
    \multirow{2}{*}{Environment} & $W_{\rm CO(2-1)}$ & $L_{\rm CO(2-1)}$ & \multirow{2}{*}{$f_{\rm CO(2-1)}$} & $W_{\rm HCN(1-0)}$ & $L_{\rm HCN(1-0)}$ & \multirow{2}{*}{$f_{\rm HCN(1-0)}$} \\ 
     & [\SI{}{\milli\kelvin\kilo\metre\per\second}] & [\SI{}{\kelvin\kilo\metre\per\second\square\kilo\parsec}] &  & [\SI{}{\milli\kelvin\kilo\metre\per\second}] & [\SI{}{\kelvin\kilo\metre\per\second\square\kilo\parsec}] &  \\ 
    \hline
    full galaxy & 3861 $\pm$ 10 & 822.2 $\pm$ 2.1 & 1471/3489 (42.2\,\%) & 168 $\pm$ 4 & 35.8 $\pm$ 0.8 & 275/3489 (7.9\,\%) \\
    centre & 82790 $\pm$ 76 & 217.3 $\pm$ 0.2 & 42/43 (97.7\,\%) & 7462 $\pm$ 36 & 19.6 $\pm$ 0.1 & 42/43 (97.7\,\%) \\ 
    bar & 8233 $\pm$ 35 & 147.2 $\pm$ 0.6 & 227/293 (77.5\,\%) & 303 $\pm$ 8 & 5.4 $\pm$ 0.1 & 70/293 (23.9\,\%) \\ 
    bar ends & 10875 $\pm$ 56 & 33.9 $\pm$ 0.2 & 50/51 (98.0\,\%) & 464 $\pm$ 15 & 1.4 $\pm$ 0.0 & 26/51 (51.0\,\%) \\ 
    spiral arms & 4577 $\pm$ 13 & 264.8 $\pm$ 0.8 & 594/948 (62.7\,\%) & 129 $\pm$ 6 & 7.5 $\pm$ 0.4 & 97/948 (10.2\,\%) \\ 
    interarm & 1178 $\pm$ 11 & 154.9 $\pm$ 1.4 & 558/2154 (25.9\,\%) & 18 $\pm$ 4 & 2.4 $\pm$ 0.5 & 40/2154 (1.9\,\%) \\
     \hline\hline
    \end{tabular}
    \end{center}
    {\bf Notes:} Columns 2 and 5 list the stacked integrated intensities of \cotwo and \hcnone for the full galaxy (first row) and respective morphological environments. The stacked spectra are shown in Fig.~\ref{fig:stacks_envir}. The corresponding \cotwo and \hcnone line luminosities are shown in columns 3 and 6. Columns 4 and 7 present the detection fractions of \cotwo and \hcnone, respectively, by environment. Shown are the number of detected ($\rm{S/N}\geq 3$) independent sight lines w.r.t. the total number of sight lines in the respective region and the corresponding detection fraction in percent in brackets.
\end{table*}

\begin{figure*}
\includegraphics[width=\textwidth]{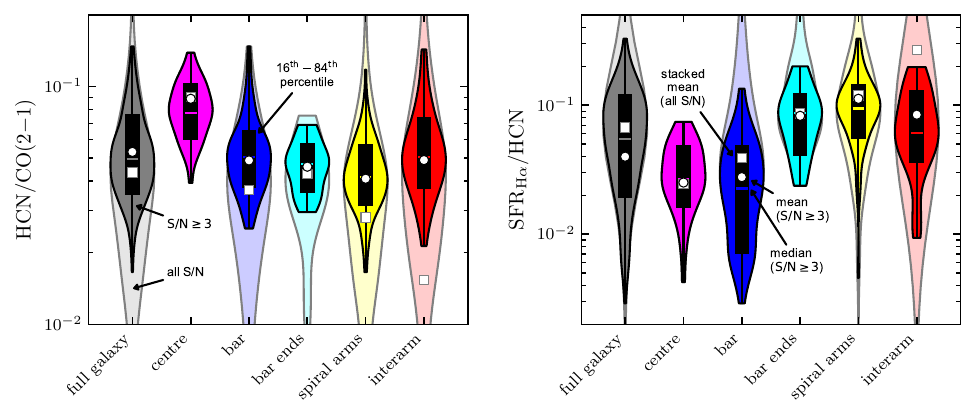}
\caption{Violin plots of dense gas spectroscopic ratios separated by environmental regions. \textit{Left:} HCN/CO line ratio, a proxy of the dense gas fraction, \fdense.
\textit{Right:} \sfrha/HCN, a proxy of the dense gas star formation efficiency, \sfedense, in units of $(\SI{}{\Msun\per\year\per\square\kilo\parsec})/(\SI{}{\Kkms})$.
The darker coloured violin areas are created from significant data ($\mathrm{S/N}\geq 3$). The lighter coloured areas represent the distribution of all S/N, this means including non-detections.
The black bar indicates the 16th to 84th percentile range of the significant data and the coloured vertical line within the bars denotes the median.
Circles show the means (sum of the ratio's numerator divided by the sum of the ratio's denominator) of the significant data.
Squares show the stacked means, which take into account all data detected in CO within the respective environment.
The values are listed in Tab.~\ref{tab:histogram_statistics}}
\label{fig:hcn_ratios_violins}
\end{figure*}

\section{Results}
\label{sec:results}

\subsection{Dense gas spectroscopic ratios across the full disc of NGC\,4321 at 260\,pc scale}
\label{sec:results:spatial_variations}
The new high-resolution, deep, wide-field HCN observations presented in this work, allow us for the first time to study variations of HCN/CO, a proxy of \fdense, and SFR/HCN, a proxy of \sfedense, across the full disc of a Milky Way-like galaxy at unprecedented resolution (\SI{260}{\parsec}) such that morphological environments can be well separated.
These data represent one of the rare deep, wide-field HCN maps of a nearby galaxy which allows analysis of 275 detected sight lines even outside of the galaxy centre as illustrated by Fig.~\ref{fig:maps_observations} (top left panel).
By environment, we detect 42 HCN sight lines in the centre, 70 in the bar, 26 in the bar ends, 97 in the spiral arms, and 40 in the interarm regions (we list the values along with stacked integrated intensities and luminosities in Tab.~\ref{tab:stacking_values}).
Figure~\ref{fig:maps_observations} shows that where CO is detected, HCN is also often detected, though HCN is found to be more concentrated in the centre, bar, bar ends and spiral arms.
To first order, \halpha and \SI{21}{\micron} emission, tracers of SFR, are spatially correlated with both CO and HCN.
The \SI{260}{\parsec} scale, resolved observations of \gal confirm the well-established linear correlation between HCN luminosity and SFR.

In Figure~\ref{fig:hcn_ratio_maps}, we show maps of HCN/CO and SFR/HCN.
The variability of these ratios provides information about how HCN correlates with CO and SFR, as discussed below.
In the following, we distinguish between five environmental regions (centre, bar, bar ends, spiral arms, interarm) introduced in Sect.~\ref{sec:methods:environments}.
The right panel of Figure~\ref{fig:hcn_ratio_maps} shows the applied environments sampled onto the same coordinate grid and overlaid with HCN contours.
Overall, the HCN emission follows the stellar mass morphology such that outside of the centre, most of the HCN emission is associated with the stellar spiral arms, whereas less HCN is found in the interarm regions.
However, there is also large amounts of (dense) molecular gas when following the bar eastward beyond the bar ends.
These regions, here depicted as interarm regions, could be interpreted as minor spiral arms or spurs between the spiral arms that harbour large amounts of molecular gas (similar to the spurs observed in M51; \citealp{Schinnerer2017}).
Though not explored here and considered part of the interarm environment, it could be interesting to study these spurs in more detail in further studies.

\subsubsection{HCN/CO variations}

Figure~\ref{fig:hcn_ratios_violins} (left panel) shows the distribution of HCN/CO values in different environments, stacked in increments of \SI{0.1}{\dex}.
We also show the mean and scatter of the detected data of the respective distributions.
Since S/N clipping systematically selects luminous HCN regions, these values will be biased towards high HCN/CO and low SFR/HCN, with the significance of the bias depending on the completeness of detections in the respective environments.
Therefore, we also show the stacked means of all sight lines across each environment (squares in Fig.~\ref{fig:hcn_ratios_violins}).
The values are listed in Tab.~\ref{tab:histogram_statistics}.

We find that HCN/CO spans roughly \SI{0.6}{\dex} when considering only detected lines of sight ($\mathrm{S/N}\geq 3$).
In agreement with previous studies \citep[e.g.][, Beslic et al. subm.]{Usero2015, Bigiel2016, Gallagher2018a, Jimenez-Donaire2019, Querejeta2019, Neumann2023a}, HCN/CO increases towards the centre of the galaxy, where it reaches values around $0.1$ (mean of $0.089\pm 0.0003$), indicating an increase of the dense gas fraction, or average gas density in centres of galaxy or/and a change in excitation conditions, e.g. optical depth, gas temperature, or abundance \citep[e.g.][]{Jimenez-Donaire2017, Eibensteiner2022}.

Throughout the disc of the galaxy (spiral arms and interarm region), HCN/CO is lower by a factors of two (mean of $\leq 0.049$ across detections, with 1-sigma scatter of $\pm \SI{0.15}{\dex}$)  compared to the centre (mean of $0.89$) and does not show trends with radius or environment (further discussed in Sect.~\ref{sec:results:radius}).
This suggests, assuming that HCN/CO is a robust tracer of density \citep{Neumann2023a}, that the density distribution of the molecular gas, which is detected in HCN, is very similar across the disc of \gal.
However, we note that when taking into account censored data, the average HCN/CO is lower by almost a factor of two in the interarm region (mean of $0.015\pm 0.003$) than in the spiral arms (mean of $0.028\pm 0.001$).

Compared to the disc of the galaxy and taking into account non-detections, we observe enhanced HCN/CO in the bar ends (mean of $0.043\pm 0.001$) pointing towards the piling up of dense molecular gas, e.g. via gas streams from the spiral arms and the bar towards the bar ends (predicted by simulations, e.g. \citealp{Renaud2015} and observed in NGC\,3627 \citealp{Beslic2021}). 
Moreover, we observe indications of a mild gradient of HCN/CO with angular offset from the spiral arm across the southern spiral arm (Fig.~\ref{fig:hcn_ratio_maps}). If taken at face value, the found HCN/CO gradient could imply a systematic density variation across the spiral arm, changing the physical conditions of the emitting gas.


\begin{table}
    \begin{center}
    \caption{HCN/CO and SFR/HCN statistics by environment.}
    \label{tab:histogram_statistics}
    \resizebox{\columnwidth}{!}{
    \begin{tabular}{ccccc}
    \hline \hline
    \multirow{2}{*}{Ratio} & \multirow{2}{*}{Environment} & (16$^{\rm th}$, 84$^{\rm th}$) perc. & mean  & stacks \\
     &  & $\mathrm{S/N}\geq 3$ & $\mathrm{S/N}\geq 3$ & all S/N \\
    \hline
    \multirow{6}{*}{$\dfrac{\rm HCN}{\cotwo}$}  & full galaxy & (0.035, 0.076) & 0.067 & $0.043\pm 0.001$ \\
      & centre      & (0.060, 0.104) & 0.089 & $0.090\pm 0.001$ \\
      & bar         & (0.037, 0.065) & 0.049 & $0.037\pm 0.001$ \\
      & bar ends    & (0.036, 0.058) & 0.046 & $0.043\pm 0.001$ \\
      & spiral arms & (0.031, 0.057) & 0.041 & $0.028\pm 0.001$ \\
      & interarm    & (0.037, 0.074) & 0.049 & $0.015\pm 0.003$ \\ 
    \hline
    \multirow{6}{*}{$\dfrac{\rm{SFR}_{\mathrm{H}\alpha}}{\rm{HCN}}$} & full galaxy & (0.019, 0.122) & 0.040 & $0.069\pm 0.002$ \\
     & centre      & (0.016, 0.049) & 0.025 & $0.025\pm 0.001$ \\
     & bar         & (0.007, 0.049) & 0.028 & $0.039\pm 0.001$ \\
     & bar ends    & (0.040, 0.124) & 0.083 & $0.086\pm 0.003$ \\
     & spiral arms & (0.055, 0.146) & 0.112 & $0.125\pm 0.006$ \\
     & interarm    & (0.036, 0.131) & 0.084 & $0.274\pm 0.057$ \\
     \hline\hline
    \end{tabular}
    }
    \end{center}
    {\bf Notes:} Statistics of spectroscopic ratios across the different environments (right panel of Fig.~\ref{fig:hcn_ratio_maps}) corresponding to the distributions shown in Fig.~\ref{fig:hcn_ratios_violins}. The third column shows the 16th and 84th percentiles of the detected measurements ($\mathrm{S/N}\geq 3$). The fourth column lists the mean of the detected data, which is computed as the sum of the numerator data over the sum of the denominator data of the spectroscopic ratio. The fifth column shows the stacks mean over all S/N along with measurement uncertainties (in dex) in the respective environment. The measurement uncertainties are computed from the rms of the emission-free channels of the stacks (Equ.~\ref{equ:emom0}) and do not include any calibration or systematic uncertainties.
\end{table}

\begin{figure*}
\centering
\includegraphics[width=\textwidth]{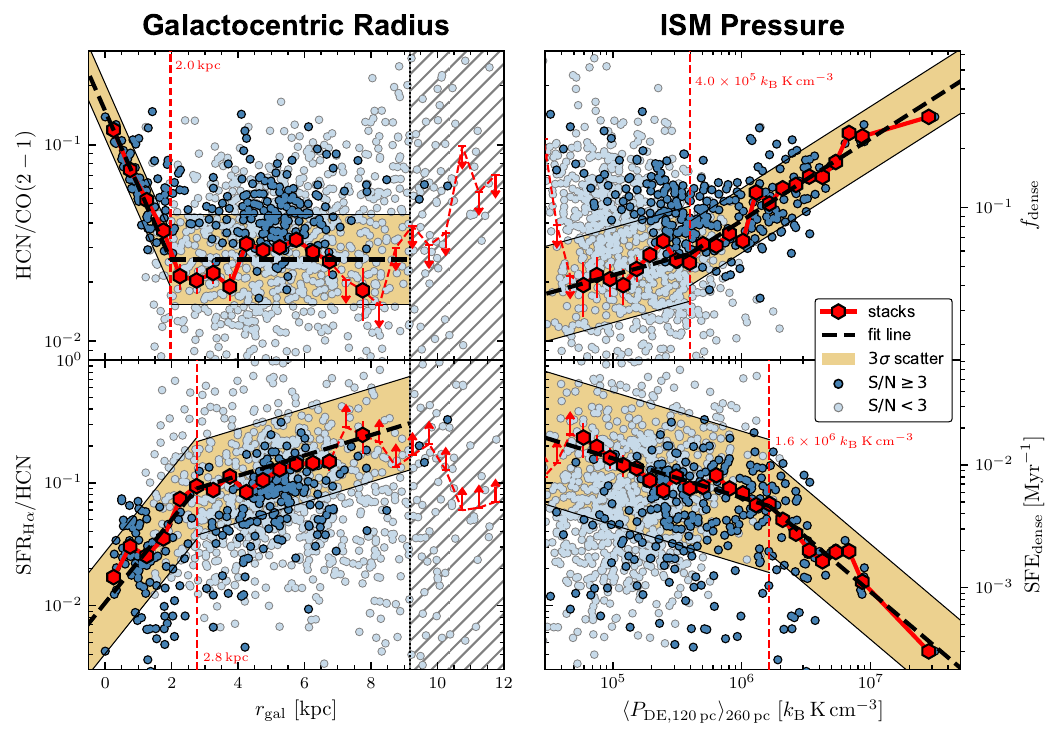}
\caption{Dense gas spectroscopic ratios as a function of galactocentric radius and environmental pressure. \textit{Top:} HCN/CO, a proxy of the dense gas fraction, \fdense, as a function of \rgal and \PDEavg. \textit{Bottom:} SFR/HCN, a proxy of the dense gas star formation efficiency, \sfedense, against \rgal and \PDEavg. Significant data, i.e. $\mathrm{S/N}\geq 3$, are shown as blue markers. Low significance data ($\mathrm{S/N}< 3$) are shown in light blue. The red hexagon markers denote significant spectral stacks taken over all data, with the bars showing the uncertainties obtained from the stacked spectra. The red arrows indicate $3\sigma$ upper limits of the HCN stacks resulting in HCN/CO upper limits and SFR/HCN lower limits. In the left panels, the hatched region ($\rgal>\SI{9.17}{\kilo\parsec}$) indicates the range where the map is not complete (compare with Fig.~\ref{fig:hcn_ratio_maps}). The vertical red dashed lines indicate the $x$-axis values separating two regions with different linear regression behaviour based on the MARS model. The dashed black lines indicate the best-fit lines resulting from the MARS model (Tab.~\ref{tab:fit_parameters}). The gold-shaded area shows the 3-sigma scatter of the detected sight-lines about the fit line.}
\label{fig:hcn_ratios_vs_radius_and_pressure}
\end{figure*}



\begin{table*}
    \begin{center}
    \caption{HCN/CO($2-1$) and SFR/HCN correlations.}
    \label{tab:fit_parameters}
    \resizebox{\textwidth}{!}{
    \begin{tabular}{ccccccc}
    \hline \hline
    x-axis & y-axis & Regime & Slope (stacks) & Slope (los) & Corr. ($p$) & Scatter \\
    \hline
    \multirow{4}{*}{\rgal} & \multirow{2}{*}{HCN/CO($2-1$)} & $\leq\SI{2.0}{\kilo\parsec}$       & -0.38 & -0.26 (0.02) & -0.86 (0.0)   & 0.14 \\ 
                           &                         & $>\SI{2.0}{\kilo\parsec}$                 & 0.00  & 0.00 (0.01)  & -0.03 (0.306) & 0.23 \\ 
                           & \multirow{2}{*}{SFR$_{\halpha}$/HCN} & $\leq\SI{2.8}{\kilo\parsec}$ & 0.34  & 0.22 (0.06)  & 0.32 (0.0)    & 0.41 \\ 
                           &                                      & $>\SI{2.8}{\kilo\parsec}$    & 0.08  & 0.08 (0.01)  & 0.41 (0.0)    & 0.38 \\ 
    \hline
    \multirow{4}{*}{\PDEavg} & \multirow{2}{*}{HCN/CO($2-1$)} & $\leq\SI{4.0e5}{\kB\kelvin\per\cubic\centi\metre}$       & 0.18  & 0.23 (0.02)  & 0.47 (0.0)    & 0.24 \\ 
                             &                         & $>\SI{4.0e5}{\kB\kelvin\per\cubic\centi\metre}$                 & 0.42  & 0.40 (0.03)  & 0.64 (0.0)    & 0.16 \\ 
                             & \multirow{2}{*}{SFR$_{\halpha}$/HCN} & $\leq\SI{1.6e6}{\kB\kelvin\per\cubic\centi\metre}$ & -0.32 & -0.53 (0.04) & -0.47 (0.0)   & 0.54 \\ 
                             &                                      & $>\SI{1.6e6}{\kB\kelvin\per\cubic\centi\metre}$    & -0.89 & -0.70 (0.22) & -0.39 (0.003) & 0.39 \\ 
     \hline\hline
    \end{tabular}
    }
    \end{center}
    {\bf Notes} -- Linear regression parameters for the respective relations and $x$-axis regimes presented in Fig.~\ref{fig:hcn_ratios_vs_radius_and_pressure}. The forth column is showing the slopes obtained from the multivariate adaptive regression spline (\texttt{MARS}) method, producing a continuous piece-wise linear regression. Columns five to seven show the linear regression parameters obtained from the individual sight-line measurements, taking into account all data in the respective regime, using the linear regression tool \texttt{LinMix}. `Corr.' is the Pearson correlation coefficient of sight-line data and `$p$' is the associated $p$-value. The scatter denotes the $3\sigma$ standard deviation of the detected sight-line data about the fit line. The cloud-scale pressure thresholds correspond to \SI{260}{\parsec}-scale beam-matched values of $\PDE\approx\SI{1.5e5}{\kB\kelvin\per\cubic\centi\metre}$ and $\PDE=\SI{6.3e5}{\kB\kelvin\per\cubic\centi\metre}$, respectively.
\end{table*}

\subsubsection{SFR/HCN variations}
Analogously to HCN/CO, we show violin plots along with mean scatter bars of SFR/HCN, a proxy of \sfedense, in the right panel of Figure~\ref{fig:hcn_ratios_violins}.
In total, the SFR/HCN values span about \SI{2}{\dex} across the detected LOSs indicating a large scatter in SFR/HCN, consistent with the cloud-to-cloud variation found in Galactic studies \citep[e.g.][]{Moore2012, Eden2012, Csengeri2016, Urquhart2021}.
Certainly, some of the scatter can be attributed to systematic variations with molecular gas conditions \citep[e.g.][]{Neumann2023a} and environment (discussed in this work).

In the inner $\sim \SI{4}{\kilo\parsec}$ of \gal, SFR/HCN appears to be spatially anti-correlated with HCN/CO (Fig.~\ref{fig:hcn_ratio_maps}), confirming kpc-scale measurements of previous studies, e.g. \citealp{Usero2015, Gallagher2018a, Jimenez-Donaire2019}.
As has been reported in several previous studies \citep[e.g.][]{Chen2015, Beslic2021, Neumann2023a}, SFR/HCN decreases towards the centre of the galaxy (mean of $0.025\pm 0.0001$ in units of $(\SI{}{\Msun\per\year\per\square\kilo\parsec})/(\SI{}{\Kkms})$) supporting the picture that HCN traces more of the bulk material in dense environments.
In addition to the centre, we find SFR/HCN to be particularly low in the bar of the galaxy (mean of $0.039\pm 0.001$) (further discussed in Sect.~\ref{sec:discussion:bar}), while it is higher by a factor of two to seven across the disc (i.e. bar ends, spiral arms and interarm regions, have means between $0.086\pm 0.003$ and $0.274\pm 0.057$).

The low SFR/HCN in the centre and bar environments can be explained in several ways.
On the one hand, the low SFR/HCN can be caused by an increase in HCN emissivity.
On the other hand, it could indicate a decrease in SFR at fixed HCN emission, i.e. an actually reduced star formation efficiency of dense gas.
Another alternative explanation put forward in previous works \citep[e.g.][]{Gallagher2018a, Jimenez-Donaire2019, Neumann2023a} is that, in these high-density, high-pressure environments, HCN is not tracing the actual overdensities anymore, but become more of a bulk molecular gas tracer.
The former can be caused by radiative trapping \citep{Shirley2015, Jimenez-Donaire2017}, lowering the effective critical density of HCN and yielding subthermally excited HCN emission \citep{Leroy2017a, Jones2023, Garcia-Rodriguez2023} or electron excitation \citep{Goldsmith2017} boosting the HCN emission.
The reduced \sfedense could be the result of a strong influence of gas dynamics on the star formation process (bar), e.g. shear, hampering the formation of stars despite the availability of dense gas (Sect.~\ref{sec:discussion:bar}).
We note that centres are much more affected by variations in conversion factors (\aCO and \aHCN) than discs, and the SFR estimator (extinction-corrected \halpha) is potentially less accurate due to increasing dust attenuation towards centres and the effects of AGN-driven \halpha emission (although this galaxy has no AGN according to \citealp{Veron2010}).
Most probably, the low SFR/HCN in the centre is a combination of an increase in gas turbulence driving HCN emission and a lower HCN-to-dense gas conversion factor.

Spiral arms, interarm regions and bar ends share a similar SFR/HCN distribution suggesting they are similarly efficiently converting dense gas into stars. 
This is contradictory to the hypothesis that bar ends are the sites of increased star formation efficiency, e.g. via cloud-cloud collision that might boost the SFE \citep{Watanabe2011, Maeda2021}.
However, we note that the aforementioned works investigate the star formation efficiency of the bulk molecular gas, traced via SFR/CO.
Therefore, their implications are likely not applying to our study of SFR/HCN, since high SFR/CO does not imply high SFR/HCN.

Overall, spiral arms and interarm regions show comparable HCN/CO and SFR/HCN distributions and means across the detected sight lines (HCN/CO of $0.041$ and $0.049$, SFR/HCN of $0.112$ and $0.084$, respectively in spiral arms and interarm regions) which demonstrates that, although spiral arms appear to show higher gas pressure and accumulate gas, they do not contain higher density clouds nor are more efficiently converting the dense gas to stars.
This result agrees with findings from Milky Way clouds \citep[e.g.][]{Dib2012, Moore2012, Eden2012, Eden2013, Eden2015, Ragan2016, Ragan2018, Rigby2019, Urquhart2021} as well as supported by high-resolution simulations of galactic-scale star formation \citep[e.g.][]{Tress2020}.

\begin{figure*}
\centering
\includegraphics[width=\textwidth]{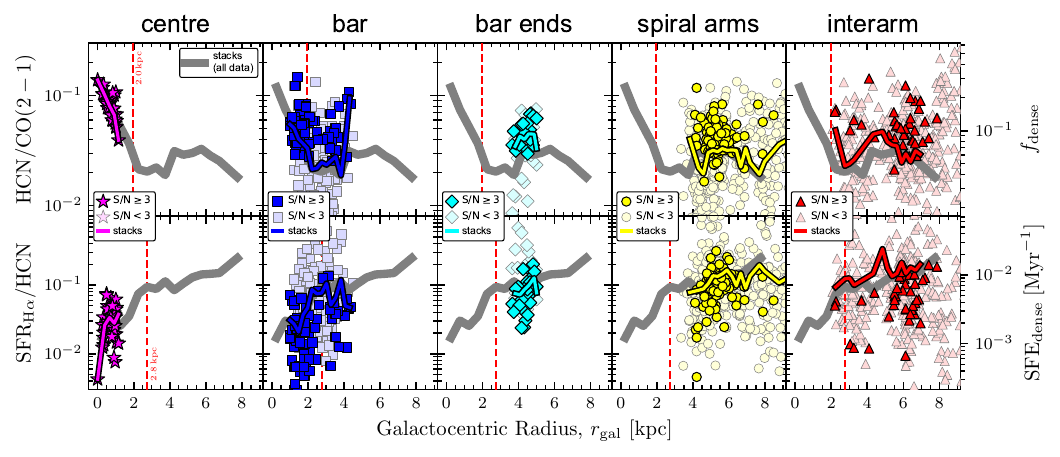}
\caption{Dense gas spectroscopic ratios vs radius in different morphological environments. Similar to Fig.~\ref{fig:hcn_ratios_vs_radius_and_pressure} (left panels), but separately for each environment (compare with right map in Fig.~\ref{fig:hcn_ratio_maps}). The lighter markers denote low-significant ($\mathrm{S/N}<3$) data. The grey solid line shows the trend of the spectral stacks as in Fig.~\ref{fig:hcn_ratios_vs_radius_and_pressure} (red markers). The coloured lines indicate the stacked measurements of the respective environments taken over all CO detected data (i.e. including HCN non-detections) in the respective environment.}
\label{fig:hcn_ratios_vs_radius_env}
\end{figure*}


\subsection{Radial trends}
\label{sec:results:radius}

Figure~\ref{fig:hcn_ratios_vs_radius_and_pressure} (left) shows the variation of HCN/CO and SFR/HCN with galactocentric radius.
To show the average trend (red markers), we stack the data in radial bins of \SI{0.5}{\kilo\parsec} width, i.e. at twice the beam size.
We then fit a piece-wise linear regression model (MARS), using the \texttt{R}-package \texttt{earth} as described in Sect.~\ref{sec:methods:fitting}.
The resulting piece-wise linear regression parameters are listed in Tab.~\ref{tab:fit_parameters}.
The MARS model finds two regimes in each of the radial correlations, which separates the relations into a central region ($\leq\SI{2.0}{\kilo\parsec}$ for HCN/CO and $\leq\SI{2.8}{\kilo\parsec}$ SFR/HCN) and a disc region (outside of the aforementioned thresholds).
The central region covers about half of the bar length, which extends out to \SI{4}{\kilo\parsec}

We note that the apparent offset between the significant data ($\rm{S/N}\geq 3$; dark blue markers) and the stacked average trends is expected in the presence of low HCN detection fraction, which is the case for most data at high \rgal or low \PDEavg. While the stacks take into account the non-detections thus recovering the true, unbiased average value, the 3-sigma clipped data are (on average) biased towards higher HCN/CO because the low HCN/CO sightline measurements tend to fall below the 3-sigma threshold, but are included in the stacked measurement.

\subsubsection{HCN/CO vs galactocentric radius}
In the inner \SI{2.0}{\kilo\parsec}, we measure a very strong (slope $m=-0.38\pm 0.02$, Pearson correlation coefficient $\rho=-0.86$, $p=\num{1.73e-51}$), tight (scatter of \SI{0.14}{\dex}) relation between HCN/CO and \rgal.
HCN/CO increases towards the centre of the galaxy where it is almost one order of magnitude higher than on average at larger \rgal agreeing with the spatial variations discussed in Section~\ref{sec:results:spatial_variations}.
Assuming HCN/CO traces density, this suggests that the fraction of dense gas is higher in the centre, consistent with resolved observations of galaxies \citep[e.g.][]{Bigiel2016, Gallagher2018a, Jimenez-Donaire2019, Beslic2021, Neumann2023a}.

Across the disc ($\rgal>\SI{2.0}{\kilo\parsec}$), HCN/CO remains constant on average ($m=0.0\pm 0.01$, $\rho=-0.14$, $p=0.306$) suggesting a more constant cloud mean density outside of galaxy centres.
However, we observe a large scatter (\SI{0.23}{\dex}) about the fit line, indicating substantial variations in HCN/CO depending on the exact location in the galaxy.
Overall, this means that outside of the centre of \gal \rgal is not a good predictor of HCN/CO at \SI{260}{\parsec} resolution.

\subsubsection{SFR/HCN vs galactocentric radius}
Similar to HCN/CO, in the central \SI{2.8}{\kilo\parsec}, SFR/HCN varies systematically with radius ($m=0.34\pm 0.02$, $\rho=0.32$, $p=\num{1.64e-05}$), though with the opposite sign ($m=0.08\pm 0.01$, $\rho=0.41$, $p=\num{1.64e-05}$).
SFR/HCN drops towards the centre ($\mathrm{SFR/HCN}\sim\num{1e-2}$) by roughly one order of magnitude with respect to the disc average value ($\mathrm{SFR/HCN}\sim\num{1e-1}$) which, taken at face value, points towards galaxy centres being less efficient in converting dense gas to stars in line with many previous works studying dense gas via HCN \citep[e.g.][Beslic et al. in prep.]{Usero2015, Bigiel2016, Gallagher2018a, Jimenez-Donaire2019, Querejeta2019, Beslic2021, Neumann2023a}.
However, we note that both HCN (due to optical depth and excitation effects) and \halpha (due to increased extinction, though here supported by additional SFR tracers; see App.~\ref{sec:appendix:sfr_tracers} for a discussion about SFR tracers in the galaxy centre) are expected to become less robust tracers of dense gas mass and SFR in galaxy centres thus mitigating any conclusions about \sfedense in these regions.

\subsection{ISM pressure relations}
\label{sec:results:pressure}

Similar to the radial trends (Sect.~\ref{sec:results:radius}), we employ the MARS tool to the stacked data in order to find regimes with different linear behaviour.
The fit results are listed in Tab.~\ref{tab:fit_parameters} and shown in the right-hand panels of Fig.~\ref{fig:hcn_ratios_vs_radius_and_pressure}.
We find a pressure threshold in both relations (HCN/CO and SFR/HCN) at $\PDEavg\approx\SI{4e5}{\kB\kelvin\per\cubic\centi\metre}$ and $\PDEavg\approx\SI{1e6}{\kB\kelvin\per\cubic\centi\metre}$ computed via Equ.~\ref{equ:PDE_avg} (the threshold value is shown as a contour overlaid on the galaxy map in the Appendix, Fig.~\ref{fig:maps_pressure_contours}; the corresponding pressure values using the alternative Equ.~\ref{equ:PDE_large} are $\PDE\approx\SI{4e5}{\kB\kelvin\per\cubic\centi\metre}$ (HCN/CO) and $\PDE\approx\SI{1e5}{\kB\kelvin\per\cubic\centi\metre}$ (SFR/HCN)).
We note that our cloud-scale \PDEavg measurements yield a factor of two to three larger values than the beam-matched \SI{260}{\parsec}-scale pressure measurements. For better comparison with previous studies that have no access to $\sim\SI{100}{\parsec}$-scale molecular gas measurements, we quote the corresponding threshold pressure values of $\PDE\approx\SIrange{1.5e5}{6.3e5}{\kB\kelvin\per\cubic\centi\metre}$, which consider the CO data convolved to \SI{260}{\parsec}-scale (opposed to the weighted average of the \SI{120}{\parsec}-scale CO measurements).

\subsubsection{HCN/CO vs pressure}
We find a strong positive correlation between the (\PDEavg-average) HCN/CO and the ISM pressure, \PDEavg, in both the high ($\rho=0.64$, $p=\num{2.84e-34}$) and low-pressure regime ($\rho=0.79$).
The correlation is steeper in the high-pressure ($m=0.42\pm 0.03$) regime compared to the low-pressure regime ($m=0.18\pm 0.03$).
However, the realtion could also be well fitted with a single linear function with $m=0.35\pm 0.02$.
Thus, the average HCN/CO increases in a roughly uniform way from $\PDEavg\approx\SIrange{5e4}{3e7}{\kB\kelvin\per\cubic\centi\metre}$ suggesting that the ISM pressure is well correlated with the average HCN/CO ($\rho=0.75$) over almost three orders of magnitude in pressure.

\subsubsection{SFR/HCN vs pressure}       
In the high-pressure regime we find a moderate negative correlation ($\rho=-0.39$, $p=0.003$,$m=-0.89\pm 0.22$) between the (\PDEavg-average) SFR/HCN and ISM pressure extending over two order of magnitude in $x$- and $y$-axis.
In the low-pressure regime the relation is significantly flatter ($m=-0.32\pm 0.04$) than in the high-pressure regime showing that across the disc, where the ISM pressure is low, SFR/HCN seems to partly decouple from the environmental pressure.
However, in both regimes, there is a significant scatter (\SI{0.39}{\dex} to \SI{0.54}{\dex}) about the average relation indicating that SFR/HCN is likely affected by other physical conditions than just the pressure or cloud properties (see App.~\ref{sec:appendix:cloud_relations}), e.g. star formation time scales or gas dynamics, where the latter could play a major role in galaxy centres and bars.

\begin{figure*}
\centering
\includegraphics[width=\textwidth]{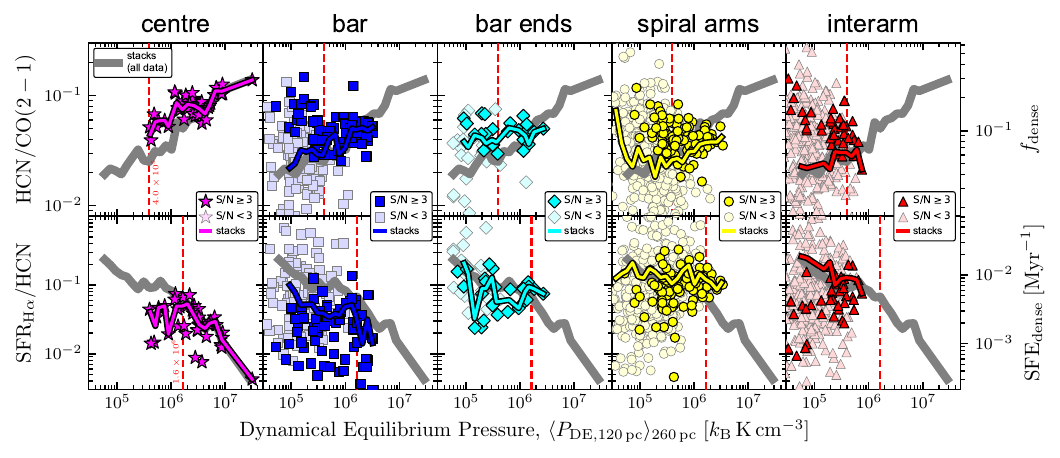}
\caption{Dense gas spectroscopic ratios vs pressure in different morphological environments. Similar to Fig.~\ref{fig:hcn_ratios_vs_radius_env}, but as a function of the dynamical equilibrium pressure.}
\label{fig:hcn_ratios_vs_pressure_env}
\end{figure*}


\subsection{Morphological environments}
\label{sec:results:environments}

In the next step, we analyse the individual morphological environments (centre, bar, bar ends, spiral arms, and interarm, as in Sect.~\ref{sec:results:spatial_variations}) in the above-discussed scaling relations.
The radial relations (Fig.~\ref{fig:hcn_ratios_vs_radius_env}) show that the centre and bar are well separated from the other environments as a function of galactocentric radius, completely dominating the strong negative (HCN/CO) and positive (SFR/HCN) trends with \rgal.
At larger radii, i.e. $\rgal\gtrsim\SI{2.5}{\kilo\parsec}$, we find several overlapping environments (bar, bar ends, spiral arms, and interam) as a function of radius.
The mean trends of the HCN/CO and SFR/HCN vs \rgal relation for each environment show an identical behaviour as the global relation shown in Sect.~\ref{sec:results:environments} and we do not find a difference between these environments.


In the HCN/CO vs \PDEavg relation (top row of Fig.~\ref{fig:hcn_ratios_vs_pressure_env}), we observe parallel trends among all environments, with the bar ends and the centre being shifted to higher HCN/CO values (following the mean trends presented as coloured lines).
Spiral arms and interarm regions show similar HCN/CO vs pressure relations suggesting that in these environments the molecular clouds have a similar mean density.

In the SFR/HCN vs \PDEavg relations (bottom row of Fig.~\ref{fig:hcn_ratios_vs_pressure_env}) the strong trend in the high-pressure regime is again dominated by the centre where SFR/HCN drops by one order of magnitude with increasing ISM pressure.
Comparing spiral arms and interarm regions, we find very similar, almost flat trends showing that spiral arms and interarm regions have similar SFR/HCN across $\SIrange{1}{2}{\dex}$ of ISM pressure and that across the disc SFR/HCN is less dependent on the ISM pressure.
In the bar ends we also find a flat trend as a function of pressure but shifted to lower SFR/HCN compared to the disc.
The bar, despite having similar HCN/CO than the disc, shows much lower SFR/HCN across the whole range of ISM pressure, more consistent with the values found in the centre.
This shows that the bar region is a peculiar environment in its star formation properties (see Sect.~\ref{sec:discussion:bar} for further discussion).

\subsection{HCN/CO as a density tracer}
\label{sec:discussion:hcn_co}

Extragalactic studies of nearby galaxies at kpc-scales \citep[e.g.][]{Gallagher2018a, Jimenez-Donaire2019}, report a positive correlation of the HCN/CO line ratio with the kpc-scale molecular gas surface density (\sigmol) as traced by the CO line intensity over more than two orders of magnitude.
These observational results are supported by theoretical works which show that HCN/CO is expected to positively correlate with the dense gas fraction and the mean gas density \citep{Leroy2017a}.
The physical interpretation put forward for explaining the strong relation between HCN/CO and \sigmol is that HCN/CO is expected to trace the density distribution of molecular clouds within the beam.
This interpretation is strongly supported by recent works \citep{Gallagher2018b, Neumann2023b} which compared the kpc-scale HCN/CO with the cloud-scale \sigmol finding a strong positive correlation.
Recently, \citet{Tafalla2023} measured the HCN/CO vs \htwo column density relation in three Solar Neighbourhood clouds, finding a similar, strong positive correlation, at least qualitatively in agreement with the extragalactic results.

\begin{figure}
\includegraphics[width=\columnwidth]{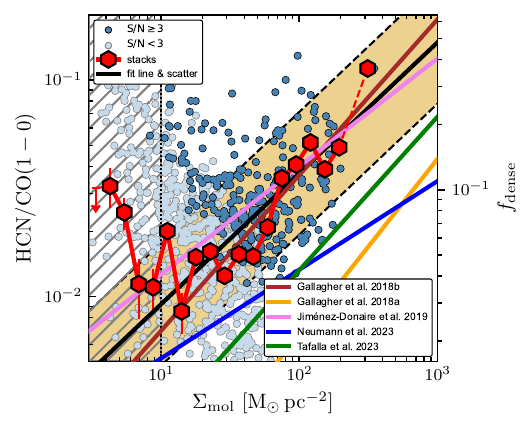}
\caption{HCN/CO vs \sigmol relation. Here, we converted the \cotwo intensities into \coone intensities using the $R_{21}$ map introduced in App.~\ref{sec:appendix:aco}. Markers show \SI{260}{\parsec} sight-line measurements from \gal, with dark blue markers denoting significant (${\rm S/N}\geq 3$) data. The fit line was obtained by linear regression using \texttt{LinMix} to the stacked measurements (red hexagons) excluding data below $\sigmol=\SI{10}{\Msun\per\square\parsec}$. The gold-shaded region shows the 1-sigma scatter of the detected sight-lines about the fit line. In addition, we show best-fit relations from literature, covering kpc-scale measurements of nearby galaxies \citep{Gallagher2018b, Gallagher2018a, Jimenez-Donaire2019, Neumann2023a}, where \citet{Gallagher2018b} and \citet{Neumann2023a} use cloud-scale CO to measure cloud-scale \sigmol and sub-pc-scale Galactic measurements of three nearby clouds \citep{Tafalla2023}.}
\label{fig:hcn_co_scaling}
\end{figure}

With the new \SI{260}{\parsec}-scale HCN observations of \gal, we can now take the next step and study the relation between HCN/CO and \sigmol at sub-kpc scales, making these results more comparable to Galactic works.
In Fig.~\ref{fig:hcn_co_scaling}, we present the relation between the \hcnone-to-\coone line ratio and \sigmol, measured at \SI{260}{\parsec} resolution.
Here, \sigmol is inferred from the \cotwo line intensity using the lower-resolution $R_{21}$ map and the surface density-metallicity based \aCO prescription as described in Sect.~\ref{sec:methods:sigmol}.
We note, that here we use the HCN/\coone, inferred from the \cotwo using the estimated $R_{21}$, instead of the HCN/\cotwo line ratio in order to better compare with literature relations.
The resulting dense gas fraction, \fdense, shown as a secondary $y$-axis, is computed using a MW-based, constant $\aCO=\SI{4.35}{\Msun\per\square\parsec}\,(\SI{}{\Kkms})^{-1}$ and a constant $\aHCN=\SI{14}{\Msun\per\square\parsec}\,(\SI{}{\Kkms})^{-1}$. Hence \fdense is assumed to be proportional to HCN/\coone (App.~\ref{sec:appendix:aco}).

\begin{table*}
    \begin{center}
    \caption{HCN/\coone vs \sigmol relations}
    \label{tab:hcn_co_relation}
    \resizebox{\textwidth}{!}{
    \begin{tabular}{ccccccc}
    \hline \hline
    $m$ (unc.) & $b$ (unc.) & $\sigma$ &\sigmol res. & HCN/CO res. & \sigmol method & Reference \\
    (1) & (2) & (3) & (4) & (5) & (6) & (7) \\
    \hline
    0.61 (0.10) & -2.64 (0.21) & 0.28 & \SI{260}{\parsec} & \SI{260}{\parsec} & beam-matched & this work\\
    0.72 (---) & -2.88 (---) & --- & $300-\SI{600}{\parsec}$ & $300-\SI{600}{\kilo\parsec}$ & beam-matched & \citet{Gallagher2018a}\\
    0.81 (0.09) & -3.80 (0.21) & --- & \SI{120}{\parsec} & \SI{650}{\parsec} & CO-weighted average & \citet{Gallagher2018b}\\
    0.50 (0.10) & -2.40 (0.20) & --- & $\sim\SI{2}{\kilo\parsec}$ & $\sim\SI{2}{\kilo\parsec}$ & beam-matched & \citet{Jimenez-Donaire2019}\\
    0.41 (0.03) & -2.70 (0.08) & 0.14 & $\sim\SI{100}{\parsec}$ & $1- \SI{2}{\kilo\parsec}$ & CO-weighted average & \citet{Neumann2023a}\\
    0.71 (0.03) & -3.30 (0.80) & --- & $\sim\SI{0.1}{\parsec}$ & $\sim\SI{0.1}{\parsec}$ & beam-matched & \citet{Tafalla2023}\\
    0.59 (0.03) & -2.64 (0.04) & 0.18 & $1- \SI{2}{\kilo\parsec}$ & $1- \SI{2}{\kilo\parsec}$ & beam-matched & Neumann, Jiménez-Donaire et al. in prep.\\    
    \hline\hline
    \end{tabular}
    }
    \end{center}
    {\bf Notes} -- Best-fit lines of HCN/CO vs \sigmol of the form shown in Equ.~\ref{equ:hcn_co_vs_sigmol}, with slope $m$ (column 1), intercept $b$ (column 2) and respective uncertainties. $\sigma$ (column 3) denotes the 1-sigma scatter of the significant data about the fit line. The lines are plotted in Fig.~\ref{fig:hcn_co_scaling}. Columns 4 and 5 list the respective $x-$ (\sigmol) and $y-$axis (HCN/CO) resolutions. ``Beam-matched'' (column 6) refers to a matched resolution of the $x-$ and $y-$axis data, and ``CO-weighted average'' denotes a CO intensity-weighted average measurement of \sigmol, adopted in \citet{Gallagher2018b} and \citet{Neumann2023a}. The differences between the physical scales and methodologies are discussed in the text.
\end{table*}


Analogous to previous sections and following Sect.~\ref{sec:methods:fitting}, we stack and fit the data to obtain the mean relation:
\begin{align}
    \log_{10}\left(\dfrac{\rm HCN}{\coone}\right) = -2.64 + 0.61\,\log_{10}\left(\dfrac{\sigmol}{\SI{}{\Msun\per\square\parsec}}\right) \,.
    \label{equ:hcn_co_vs_sigmol}
\end{align}
We list the relation parameters along with uncertainties and relations from the literature in Tab.~\ref{tab:hcn_co_relation}.
We note that we exclude data below $\sigmol<\SI{10}{\Msun\per\square\parsec}$ (shaded region in Fig.~\ref{fig:hcn_co_scaling}) from the fit, because at lower surface densities the trend does not seem to continue in the same manner.
This could indicate that at low surface densities, HCN/CO does not increase with \sigmol anymore.
\citet{Santa-Maria2023} argued that this could be due to HCN being excited in hot, low-surface density regions.
However, due to a lack of sensitivity below $\sigmol=\SI{10}{\Msun\per\square\parsec}$, we can not test this hypothesis with our data. 
We stress hat such a trend can be the result of low completeness thus reflecting the biased-high average HCN/CO if either HCN and CO are clipped ($\rm{S/N}\geq 3$) or if the $x$-axis is not complete.
First and foremost, at $\sigmol>\SI{10}{\Msun\per\square\parsec}$, we find a strong positive correlation between HCN/CO and \sigmol which agrees well with much of the prior literature.
Though, the scatter in the individual \SI{260}{\parsec} sight-line measurements is twice as large (\SI{0.28}{\dex}) as the scatter at kpc-scales (\SI{0.14}{\dex}).
The larger scatter at smaller scales indicates strong cloud-to-cloud variations in qualitative agreement with Galactic studies finding large \fdense variations \citep[e.g.][]{Moore2012, Eden2012, Csengeri2016, Urquhart2021, Tafalla2023}.
One explanation for the increased scatter at smaller scales can be cloud evolution effects, leading to changes in the HCN/CO line ratio over the life cycle of molecular clouds, which can only be resolved at smaller scales \citep[e.g.][]{Kruijssen2014a}.
\citet{Tafalla2023} suggest that some of the variations are caused by gas temperature variations between clouds, affecting the HCN excitation. 
In addition, HCN (and CO) emissivity can be affected by optical depth \citep{Shirley2015, Leroy2017a, Jimenez-Donaire2017, Jones2023, Garcia-Rodriguez2023} and electron excitation \citep{Goldsmith2017}, further driving the scatter about the relation.
Certainly, in-depth investigations of HCN at higher resolution in nearby galaxies are needed to understand what is driving HCN/CO at fixed surface density.

Comparing with previous literature findings, the exact relations vary significantly between different studies.
The reported slopes span values from 0.41 over 0.61 (this work) to 0.81. 
Neumann, Jiménez-Donaire et al. in prep. combine measurements from EMPIRE \citep{Jimenez-Donaire2019} and ALMOND \citep{Neumann2023a} and match methodologies to obtain an updated, more robust constraint on the HCN/CO vs \sigmol relation at kpc-scales, yielding a slope of 0.59 consistent with the slope found for \gal in this work.
Some of the differences between studies can at least partially be attributed to different methodologies.
For instance, the adopted \aCO prescription significantly affects the measured relation.
For \gal, using a constant \aCO yields a shallower relation (slope of 0.48) compared to 0.61 with a varying \aCO using the prescription described by Equ.~\ref{equ:aco}.
Moreover, the physical scales observed can significantly affect slopes (\citealp{Gallagher2018b} found a slope of 0.41 at \SI{2.8}{\kilo\parsec}-scale opposed to 0.81 at \SI{650}{\parsec}-scale).
In addition, using \cotwo instead of \coone will affect the slope if a constant $R_{21}$ is used to convert \cotwo to \coone, making the slope flatter than a native \coone measurement, since $R_{21}$ negatively correlates with \sigmol \citep{Leroy2022}.
For these reasons, comparisons between different studies have to be taken with care.
Nevertheless, we want to stress that, independent of methodologies, HCN/CO, at least qualitatively, robustly traces the average molecular gas density from sub-pc to kpc scales.

Even if tracers and methods are matched, the resolution is expected to affect the observed relation if the emission is not beam-filling. 
EMPIRE \citep{Jimenez-Donaire2019} and ALMOND \citep{Neumann2023a} both studied kpc-scale HCN/CO as a function of \sigmol.
However, ALMOND used (intensity-weighted) cloud-scale (\SI{150}{\parsec}) \sigmol, while EMPIRE used matched-resolution kpc-scale \sigmol.
Comparing the HCN/CO vs \sigmol from EMPIRE and ALMOND, we find a $\sim\SI{1}{\dex}$ shift of the relation towards higher \sigmol at smaller scales.
These results suggest that the CO filling factor is lower by a factor of $\sim 10$ at $\sim\SI{1}{\kilo\parsec}$ compared to $\sim\SI{100}{\parsec}$.
Furthermore, increasing the resolution of the HCN measurements, i.e. going to smaller-scale HCN/CO measurements, appears to make the relations steeper (see above) and shifted towards higher HCN/CO, suggesting that HCN is clumpier and/or tracing denser gas than CO.

HCN abundance is expected to vary with metallicity due to the strong decrease of nitrogen-bearing species (e.g. HCN) with decreasing metallicity opposed to oxygen-bearing species (e.g. HCO$^+$) \citep[e.g.][]{Braine2017, Braine2023}. However, across \gal the metallicity varies by only $\sim\SI{0.1}{\dex}$ (see bottom panel of Fig.~\ref{fig:maps_conversion_factors}). Therefore, abundance and optical depth variations connected to metallicity changes are expected to play only a minor role in affecting the HCN emissivity. In the appendix (Fig.~\ref{fig:metallicity_trends}), we show how the HCN-to-HCO$^+$ line ratio varies with metallicity across \gal, finding almost no dependence of HCN/HCO$^+$ on metallicity, supporting the aforementioned statement that HCN is yielding similar results as other dense gas tracers. To further support this statement, we investigated the same scaling relations that are shown in Fig.~\ref{fig:hcn_ratios_vs_radius_and_pressure}, replacing HCN by HCO$^+$ as a dense gas tracer, yielding the same trends with \rgal and \PDEavg. The average relations agree within \SI{10}{\percent} with the HCN results except for the central $\sim\SI{1}{\kilo\parsec}$, where HCN is about \SI{30}{\percent} brighter than the average line ratio value of $\rm{HCN/HCO}^+\approx 1.3$.

We note that the trend expressed by Equ.~\ref{equ:hcn_co_vs_sigmol} could be driven by the centre, where we find the strongest systematic variations in HCN/CO.
If HCN/CO is interpreted as \fdense, we expect the highest centre-to-disc variations in the \aCO and \aHCN conversion factors in the centre caused e.g. by strong variations in optical depth and excitation temperature, which could additional affect the correlation between \fdense and \sigmol.
We checked that both the centre and the disc show a significant positive correlation between HCN/CO and \sigmol with slopes of 0.25 (centre) and 0.35 (disc), which shows that even when the centre is excluded there is a clear dependence of HCN/CO on \sigmol.
However, the relations in the individual environments are much flatter compared to the overall trend (slope of 0.61) and offset by about \SI{0.4}{\dex}, which indicates that the overall HCN/CO vs \sigmol trend might be enhanced by a centre-to-disc dichotomy.

\section{Discussion}
\label{sec:discussion}

\subsection{Pressure threshold for dense gas and star formation}
Over the last decade, resolved kpc-scale observations of nearby galaxies have found a systematic correlation between high-critical density tracer ratios, i.e. HCN/CO and SFR/HCN, and the environmental pressure in the ISM disc \citep[e.g.][]{Gallagher2018a, Gallagher2018b, Jimenez-Donaire2019, Neumann2023a}.
These results are qualitatively supported by observations of the Milky Way's CMZ, where the star formation efficiency of the (dense) molecular gas is low \citep{Longmore2013, Kruijssen2014b, Henshaw2023}.
These results contrast with Solar neighbourhood results that tend to find a constant SFR/HCN \citep[e.g.][]{Heiderman2010, Lada2010, Lada2012, Evans2014}.
However, these Solar Neighbourhood observations probe a much lower ISM pressure environment ($\PDE\lesssim\SI{2e4}{\kB\kelvin\per\cubic\centi\metre}$) than typical extragalactic works.
The apparent tension between the Galactic and extragalactic works could, however, be solved if there existed a pressure threshold above which cloud properties and hence the observed spectroscopic ratios depend on pressure.
Below this threshold, the clouds would be able to decouple from the environment and show universal behaviour in converting the dense gas to stars, a concept also put forward by theoretical works \citep[e.g.][]{Krumholz2007, Ostriker2010, Krumholz2012}.
\citet{Gallagher2018a} suggests that this pressure threshold could be at $\PDE\approx\SI{2e5}{\kB\kelvin\per\cubic\centi\metre}$, which is similar to the internal pressure of a typical GMC with $\sigmol=\SI{100}{\Msun\per\square\parsec}$.
We note that our formulation of the ISM pressure includes both the environment (gas and stellar mass) as well as cloud-scale molecular gas mass leading to a factor of 2--3 higher \PDE values compared to the purely kpc-scale environmental \PDE \citep{Sun2020a}.
Therefore, the exact value of the pressure threshold is likely to vary with the resolution at which the pressure is measured.

With the new wide-field, deep HCN observations of \gal presented in this work, we can now for the first time explore the low-pressure environment ($\PDE\approx\SI{1e5}{\kB\kelvin\per\cubic\centi\metre}$) at \SI{260}{\parsec} scales in a Milky Way-like galaxy and address the question: ``Is there a pressure threshold for dense gas and star formation?''
In Figure~\ref{fig:hcn_ratios_vs_radius_and_pressure} (right panels), we determined two pressure regimes in each of the relations based on the change in the behaviour of the mean trends using the methodology described in Sect.~\ref{sec:methods:fitting}.
Focusing on the SFR/HCN vs \PDEavg relation, we find a clear negative correlation at high pressures which significantly flattens in the low-pressure regime (especially evident in the mean trends of the individual environments), with the threshold being at $\PDEavg_{\rm threshold}=\SI{1.6e6}{\kB\kelvin\per\cubic\centi\metre}$.
Thus, our results might support the pressure threshold hypothesis laid out above (slope changes by \SI{30}{\percent}), though finding a threshold that is one order of magnitude higher than the value inferred from simple theoretical considerations ($\PDE\approx\SI{2e5}{\kB\kelvin\per\cubic\centi\metre}$).
We note, however, that the measured pressure estimates depend strongly on the scales at which they are measured.
\citet{Sun2020a} show that larger physical scales ($\sim\SI{1}{\kilo\parsec}$) can lead to lower pressure estimates due to averaging out GMC-scale ($\sim\SI{100}{\parsec}$) variations of the molecular gas.
Therefore, even higher resolution observations ($\lesssim\SI{100}{\parsec}$) are needed to obtain robust quantitative pressure estimates comparable with Solar Neighbourhood measurements.

In the appendix, we also present the relations between the spectroscopic ratios and the \SI{120}{\parsec}-scale molecular gas properties (Fig.~\ref{fig:hcn_ratios_vs_cloud_props}) representing the self-gravity term on the ISM pressure as well as the relation with stellar mass surface density (Fig.~\ref{fig:hcn_ratios_vs_stars}) representing the environment term. 
We find that a threshold-like behaviour is only seen in the cloud property relation, where SFR/HCN becomes almost constant in the low-\sigmol, low-\vdis regime.
In contrast, SFR/HCN shows a monotonic negative relation with \sigstar suggesting that clouds are always connected to the environmental pressure, but in the low-pressure environment the amount of dense gas is converted into stars in a uniform way independent of the cloud-scale properties.

\subsection{Normal star formation efficiency in bar ends}
\label{sec:discussion:bar_ends}

Observations \citep[e.g.][]{Kenney1991, Harada2019, Sormani2019, Yu2022a, Yu2022b} and simulations \citep[e.g.][]{Sormani2018} show that gas inflow from the spiral arms and gas outflow from the bar can feed the bar ends with molecular gas.
As a consequence, the bar ends are the principal site for cloud-cloud collisions, which are thought to either boost \citep[e.g.][]{Habe1992, Benjamin2005, Lopez-Corredoira2007, Furukawa2009, Ohama2010, Fukui2014, Renaud2015, Fukui2016, Fukui2018, Torii2017, Sormani2020} or lower \citep[e.g.][]{Fujimoto2020} the formation of stars, depending on the relative speed of the colliding clouds \citep[e.g.][]{Takahira2014}.
his raises the question: ``Do bar ends boost or suppress star formation?''
In accordance with the picture that bar ends are fed by gas flows, we observe relatively bright HCN and CO emission, implying an accumulation of (dense) molecular gas in the bar ends.
However, we do not find an increased \sfedense, traced by SFR/HCN, compared to the rest of the disc (spiral arms and interarm), suggesting that collisions might enhance density but not necessarily lead to different processes in the dense gas.
We discuss the bar environment in the following subsection (Sect.~\ref{sec:discussion:bar}).

\subsection{Star formation suppression in the bar}
\label{sec:discussion:bar}

Galactic bars are the sites of strong shear and gas streaming motions, which can potentially affect the (density) structure of molecular clouds and their ability to form stars \citep[e.g.][]{Athanassoula1992, Emsellem2015, Sormani2018}.
\citet{Diaz-Garcia2021} find that bar strength can affect quenching, suggesting that bars are loci of suppressed star formation.
However, \citet{Fraser-McKelvie2020} also find evidence for increased star formation along bars.
\citet{Maeda2023} propose that the star formation efficiency of molecular gas (SFR/CO; \sfemol) in nearby spiral galaxies is systematically suppressed in bars.
Here, we go a step further and study the denser molecular gas (traced by HCN) that is more tightly related to SFR and its star formation efficiency, \sfedense.
On the one hand, we find very similar HCN/CO, tracing \fdense, in the bar as well as throughout the disc.
Additionally, we observe the same average trend with ISM pressure as in all other environments suggesting that bars contain clouds with similar mean density than galaxy discs.
On the other hand, the agreement with disc trends changes for the dense gas star formation efficiency (\sfedense traced by SFR/HCN), which is much lower than in the disc.
The systematically lower SFR/HCN in the bar becomes even more evident in the relation with pressure, where the average trend of the bar is $\approx\SI{0.5}{\dex}$ lower compared to the other environments.
This suggests that the bar of \gal is indeed much less efficient in converting dense molecular gas into stars despite the presence of overdense gas.

One explanation for the low \sfedense could be shearing motions inside the bar that are solenoidal in nature, lowering the star formation efficiency \citep[e.g.][]{Federrath2016} or high-speed cloud-cloud collisions \citep[e.g.][]{Fujimoto2020}.
Recent simulations \citep[e.g.][]{Sormani2018} and observations \citep{Wallace2022} suggest that gas dynamics in the bar are more dominated by streaming motions from the bar ends towards the galactic centre.
These streaming motions can result in deformation and stretching of the molecular clouds in the bar leading to elongated, destructed molecular clouds, which might counteract the gravitational collapse hence quenching star formation.

\section{Conclusions}
\label{sec:conclusions}

We present new, deep, wide-field \hcnone dense molecular gas observations at scales of \SI{260}{\parsec} of the nearby spiral galaxy \gal.
In combination with recent, high-resolution ($\sim\SI{1}{\arcsec}\sim\SI{100}{\parsec}$) observations of CO (PHANGS--ALMA, \citealp{Leroy2021b}), tracing the bulk molecular gas and \halpha observations (PHANGS--MUSE; \citealp{Emsellem2022}) tracing SFR (supported by \SI{21}{\micron} observations from PHANGS--JWST \citep{Lee2023}, we are able to study for the first time dense gas spectroscopic ratios (HCN/CO, SFR/HCN) in many individual sight lines and environments expanding into the low-pressure regime that is similar to the Solar neighbourhood environment.
We use morphological masks based on the stellar mass content \citep{Querejeta2021} to distinguish between different environmental regions that have different structural and dynamical properties potentially affecting the properties of molecular clouds and their ability to form stars.
We study how HCN/CO, a proxy of the dense gas fraction (\fdense), and SFR/HCN, a proxy of the dense gas star formation efficiency (\sfedense), vary between different galactic environments and depend on the ISM pressure.
Our key findings are:

\begin{enumerate}

    \item HCN/CO increases and SFR/HCN decreases towards the centre of the galaxy, and are roughly flat across the galactic disc.
    This suggests that galaxy centres have denser molecular clouds, but those are less efficiently converted into stars than in the disc.
    These global trends are consistent with previous results from kpc-scale surveys, but our superior resolution allows us to analyse the role of the environment in more detail. 
    Distinguishing between environmental regions (centre, bar, bar ends, spiral arms, and interarm), we find HCN/CO to be significantly higher in the centre, while SFR/HCN is lower in both the centre and the bar.
    This shows that the star formation process is roughly universal across the disc of \gal. 
    In particular, we find no significant difference of HCN/CO and SFR/HCN between the spiral arms and interarm regions, but star formation from the dense gas is significantly less efficient in the bar and centre.
    The strong trends towards the centre of \gal suggest either that clouds couple strongly to the surrounding environment or that HCN is tracing more of the bulk molecular gas that is less efficiently converted into stars, the latter being in agreement with predictions from gravo-turbulent cloud models.
    
    \item The mean dense gas spectroscopic trends in the disc ($\rgal\gtrsim\SI{2}{\kilo\parsec}$) of \gal are very similar among different environments (Fig.~\ref{fig:hcn_ratios_vs_radius_env}, Fig.~\ref{fig:hcn_ratios_vs_pressure_env}). 
    However, we observe a significant scatter of \SI{0.14}{\dex} to \SI{0.54}{\dex} in all relations at fixed radius or fixed pressure.
    This indicates that, although on average similar, cloud properties and their environment vary more significantly at smaller scales and are potentially strongly affected by local, cloud-scale ($\lesssim\SI{100}{\parsec}$) physics and local excitation conditions (e.g. gas temperature, optical depth).
    
    \item We identify a pressure threshold for dense gas and star formation at $\PDEavg_{\rm threshold}\approx\SI{1e6}{\kB\kelvin\per\cubic\centi\metre}$ (Equ.~\ref{equ:PDE_avg}; Fig.~\ref{fig:hcn_ratios_vs_radius_and_pressure}) corresponding to \SI{260}{\parsec}-scale $\PDE\approx\SI{4e5}{\kB\kelvin\per\cubic\centi\metre}$ (Equ.~\ref{equ:PDE_large}). While the relation between pressure and HCN/CO can also be well described by a single relation covering both regimes, the relation with SFR/HCN significantly flattens in the low-pressure regime. This supports the idea that there is a pressure threshold below which the star formation process in molecular clouds becomes less dependent on the environment as seen in Galactic measurements of molecular clouds. Thus, our findings hint towards resolving the tension between Galactic and extragalactic studies of dense gas and star formation.
    
    \item The bar of \gal is showing significantly lower SFR/HCN than the disc and is systematically shifted to lower SFR/HCN in relation to the ISM pressure.
    This is a strong indication that the star formation in the bar is suppressed by shear or streaming motion that prevents the gravitational collapse and thus star formation.

    \item We find a strong, positive correlation between HCN/CO and \sigmol (Fig.~\ref{fig:hcn_co_scaling}) with a slope of 0.61, using a varying \aCO to compute \sigmol (for comparison, a constant \aCO yields a slope of 0.48). Our \SI{260}{\parsec}-scale results are in agreement with many previous studies from sub-pc (Galactic) to kpc-scales (extragalactic) and support the use of HCN/CO as a powerful tracer of molecular cloud average density. We emphasise that the exact relation depends on methodology and that scatter increases at smaller scales. We find a scale-dependence of the relation likely connected to the beam filling factors of CO and HCN, indicating that HCN emission is tracing denser gas than CO and originating from smaller than \SI{260}{\parsec}-scale regions.
    
\end{enumerate}

These findings present the next step in connecting extragalactic and Galactic studies of dense gas and star formation.
Overall, our results motivate the presence of a pressure threshold for dense gas and star formation and advertise the potential to link Galactic and extragalactic works.
However, we are still not able to resolve individual GMCs ($\lesssim\SI{100}{\parsec}$) in galaxies beyond the local group and explore conditions similar to the Solar neighbourhood ($\PDE\approx\SI{2e4}{\kB\kelvin\per\cubic\centi\metre}$).
Even deeper observations of dense gas tracers are needed to find better constraints on spectroscopic dense gas ratios in the Solar neighbour-like low-pressure environment.
Moreover, we show that across the disc SFR tracers yield similar results, but towards the centre they can differ a lot, hence requiring more in-depth studies of galaxy centres to infer robust prescriptions of dense gas and star formation in these extreme environments.
Further, this work investigates only a single galaxy.
Ultimately, we need similar dense gas studies in a large sample of galaxies (as done in PHANGS) to study, e.g. the effect of bar dynamics on the star formation efficiency, or the pressure threshold hypothesis in a statistically meaningful sample of galaxies.
Besides, the scale-dependence of the HCN/CO vs. \sigmol relation requires a more detailed study, particularly addressing the CO and HCN filling factors at sub-\SI{100}{\parsec} scales.

\begin{acknowledgements}
LN acknowledges funding from the Deutsche Forschungsgemeinschaft (DFG, German Research Foundation) - 516405419.
AKL gratefully acknowledges support by grants 1653300 and 2205628 from the National Science Foundation, by award JWST-GO-02107.009-A, and by a Humboldt Research Award from the Alexander von Humboldt Foundation. The work of AKL is partially supported by the National Science Foundation under Grants No. 1615105, 1615109, and 1653300.
AU acknowledges support from the Spanish grants PID2019-108765GB-I00, funded by MCIN/AEI/10.13039/501100011033, and PID2022-138560NB-I00, funded by MCIN/AEI/10.13039/501100011033/FEDER, EU.
ER acknowledges the support of the Natural Sciences and Engineering Research Council of Canada (NSERC), funding reference number RGPIN-2022-03499.
MB acknowledges support by the ANID BASAL project FB210003 and by the French government through the France 2030 investment plan managed by the National Research Agency (ANR), as part of the Initiative of Excellence of Université Côte d’Azur under reference number ANR-15-IDEX-01.
MC gratefully acknowledges funding from the DFG through an Emmy Noether Research Group (grant number CH2137/1-1).
COOL Research DAO is a Decentralized Autonomous Organization supporting research in astrophysics aimed at uncovering our cosmic origins.
KG is supported by the Australian Research Council through the Discovery Early Career Researcher Award (DECRA) Fellowship (project number DE220100766) funded by the Australian Government. KG is supported by the Australian Research Council Centre of Excellence for All Sky Astrophysics in 3 Dimensions (ASTRO~3D), through project number CE170100013.
JDH gratefully acknowledges financial support from the Royal Society (University Research Fellowship; URF/R1/221620).
HAP acknowledges support by the National Science and Technology Council of Taiwan under grant 110-2112-M-032-020-MY3.
MQ acknowledges support from the Spanish grant PID2019-106027GA-C44, funded by MCIN/AEI/10.13039/501100011033.
TS acknowledges funding from the European Research Council (ERC) under the European Union’s Horizon 2020 research and innovation programme (grant agreement No. 694343).
ES acknowledges funding from the European Research Council (ERC) under the European Union’s Horizon 2020 research and innovation programme (grant agreement No. 694343).
SKS acknowledges financial support from the German Research Foundation (DFG) via Sino-German research grant SCHI 536/11-1.
Y-HT acknowledges funding support from NRAO Student Observing Support Grant SOSPADA-012 and from the National Science Foundation (NSF) under grant No. 2108081.
TGW acknowledges funding from the European Research Council (ERC) under the European Union’s Horizon 2020 research and innovation programme (grant agreement No. 694343).

This paper makes use of the following ALMA data \\
\noindent ADS/JAO.ALMA\#2011.0.00004.SV, \linebreak 
\noindent ADS/JAO.ALMA\#2015.1.00956.S, \linebreak 
\noindent ADS/JAO.ALMA\#2017.1.00815.S. \linebreak 
ALMA is a partnership of ESO (representing its member states), NSF (USA), and NINS (Japan), together with NRC (Canada), NSC and ASIAA (Taiwan), and KASI (Republic of Korea), in cooperation with the Republic of Chile. The Joint ALMA Observatory is operated by ESO, AUI/NRAO, and NAOJ. The National Radio Astronomy Observatory (NRAO) is a facility of the National Science Foundation operated under cooperative agreement by Associated Universities, Inc.

\end{acknowledgements}

\section*{Data availability}
The data used within this paper will be provided on reasonable request to the corresponding author.

\bibliographystyle{aa}
\bibliography{references.bib}


\begin{appendix} 

\section{Ancillary data}
\label{sec:appendix:ancillary_data}

\subsection{CO-to-\htwo conversion factor -- \aCO}
\label{sec:appendix:aco}

In this work, we employ two different conversion factors to convert the \cotwo line intensity into a molecular gas surface density as described in Sect.~\ref{sec:methods:sigmol}.

\subsubsection{Constant conversion factor}
\label{sec:methods:sigmol_const_aco}
As a first step, we study variations of the HCN-to-CO line ratio, which is a proxy for the dense gas fraction, adopting constant mass-to-light ratios (Sect.~\ref{sec:methods:fdense}).
We adopt a Milky-Way average CO-to-\htwo conversion factor of $\aCO=\SI{4.35}{\Msun\per\square\parsec}\,(\SI{}{\Kkms})^{-1}$ that is uncertain by a factor of two \citep{Bolatto2013}.
To convert the \cotwo into a \coone line intensity, we use a \cotwo-to-\coone line ratio of $R_{21}=0.65\pm 0.17$ \citep{denBrok2021, Leroy2022}.

\subsubsection{Spatially varying conversion factor}
\label{sec:methods:sigmol_varying_aco}
As a second step, we adopt varying conversion factors to obtain the most accurate estimation of the molecular gas surface density given the current knowledge about spatial variations of \aCO and $R_{21}$ in nearby galaxies at $\sim\SI{100}{\parsec}$ scales.
This prescription enters the estimation of the dynamical equilibrium pressure (Sect.~\ref{sec:methods:pde}).
We convert the \cotwo moment-0 (\intCO) into a molecular gas surface density (\sigmol) in a two-step process: First, we convert the \cotwo into a \coone line intensity by applying a spatially varying \cotwo-to-\coone line ratio ($R_{21}$) map. 
This makes use of archival \coone observations of \gal with ALMA at \ang{;;4} resolution. 
Hence we can compute $R_{21}$ at $\ang{;;4}\sim\SI{300}{\parsec}$ scale and infer \coone from the measured \cotwo at $\ang{;;1.67}\sim\SI{120}{\parsec}$ resolution:
\begin{equation}
    W_{\coone}^{\ang{;;1.67}} \approx W_{\cotwo}^{\ang{;;1.67}} / R_{21}^{\ang{;;4}}
\end{equation}
Certainly, we expect small-scale variations of $R_{21}$ at $<\SI{300}{\parsec}$ scales, but using a \SI{300}{\parsec}-smoothed $R_{21}$ still provides more accurate estimates of \coone and hence \sigmol than adopting a constant $R_{21}$.
Here, we measure a median $R_{21}=0.54$ and a scatter of \SI{0.16}{\dex}.
We show a map of $R_{21}$ in Fig.~\ref{fig:maps_conversion_factors}.

Next, we convert the inferred \coone moment-0 map ($W_{\coone}$) into \sigmol by applying the prescription described by Equation~31 from \citet{Bolatto2013}:
\begin{equation}
    \left(\dfrac{\alpha_{\coone}}{\SI{}{\Msun\per\square\parsec}\,(\SI{}{\Kkms})^{-1}}\right) = 2.9\, \exp\left(\dfrac{0.4}{Z^{\prime}\,\Sigma_{\rm GMC}^{100}}\right) \left(\dfrac{\Sigma_{\rm total}}{\SI{100}{\Msun\per\square\parsec}}\right)^{-\gamma}
    \label{equ:aco}
\end{equation}
following the iterative approach presented in \citet{Sun2022}. $Z^{\prime}$ is the local metallicity in units of Solar metallicities, and $\gamma$ is 0.5 if $\Sigma_{\rm total}>\SI{100}{\Msun\per\square\parsec}$ and 0 otherwise.
The local metallicity is estimated from optical line measurements taken by MUSE applying a Gaussian Process Regression (GPR) to model the 2D metallicity distribution \citep{Williams2022}.
Here, $\Sigma_{\rm GMC}^{100}$ is fixed at unity, i.e. we adopt a fiducial GMC surface density of \SI{100}{\Msun\per\square\parsec} and the total surface density ($\Sigma_{\rm total}$) includes molecular gas (CO), atomic gas (\hone, Sect.~\ref{sec:appendix:hi}), and stellar mass (\SI{3.6}{\micron}, Sect.~\ref{sec:appendix:mstar}).
Since \aCO is implicitly needed to compute $\Sigma_{\rm total}$, Equation \ref{equ:aco} must be solved iteratively.
We show a map of the adopted \aCO map in Fig.~\ref{fig:maps_conversion_factors}. 
The median \aCO value is $\SI{3.3}{\Msun\per\square\parsec}\,(\SI{}{\Kkms})^{-1}$ and thus by a factor of 1.3 lower than the MW based \aCO listed above. 
For this galaxy, \aCO systematically decreases by a factor of $\sim 3$ from the disc over the spiral arms towards the centre of \gal, with a scatter of \SI{0.10}{\dex}.
Overall, the inferred total \htwo mass of the varying \aCO across the FOV used in this work is $M_{\htwo}=\SI{1.6e9}{\Msun}$, which is lower by a factor of 0.57 than the value obtained via a constant MW-based \aCO, yielding $M_{\htwo}=\SI{2.8e9}{\Msun}$.

We note that \citet{Teng2023} inferred cloud-scale \aCO values for the central 2\,kpc of \gal using a multi-line modelling approach based on cloud-scale multi-line, multi-transition CO isotopologues.
Despite the existing robust \aCO measurement, we employ the surface density-metallicity-based calibration from \citet{Bolatto2013} to obtain continuous \aCO values across the full disc of \gal and note that the measurements from \citet{Teng2023} are on average \SI{0.20}{\dex} smaller than the values used here, but consistent at the 1-sigma level within \SI{0.21}{\dex}.

\subsection{Atomic gas -- \hone 21-cm}
\label{sec:appendix:hi}
We utilise \hone 21-cm line observations to measure the atomic gas surface density. 
The \hone data are from VLA observations associated with HERACLES \citep{Leroy2009} mapping several nearby galaxies in \hone at $\sim\ang{;;20}$ resolution. 
We convert the integrated intensity of \hone ($W_{\rm 21cm}$) into atomic gas surface density via \citep{Walter2008}:
\begin{equation}
   \left(\dfrac{\sigatom}{\SI{}{\Msun\per\square\parsec}}\right) = \num{1.97e-2} \left(\dfrac{W_{\rm 21cm}}{\SI{}{\Kkms}}\right) \cos(i),
\end{equation}
where $\cos(i)$ is accounting for the inclination of the galaxy ($i=\SI{38.5}{\degree}$).

\subsection{Stellar mass -- \SI{3.6}{\micro\metre}}
\label{sec:appendix:mstar}
We use the stellar mass surface density map from \citet{Querejeta2015}, who use \SI{3.6}{\micro\metre} observations from the \textit{Spitzer} Survey of Stellar Structure in Galaxies \citep[S$^4$G;][]{Sheth2010}. 
The \SI{3.6}{\micro\metre} maps are corrected for dust attenuation using an ``Independent Component Analysis'' (ICA) method that separates stars and dust on a pixel-to-pixel basis (for more details, see \citealp{Querejeta2015})
We convert the attenuation-corrected \SI{3.6}{\micro\metre} map into stellar mass surface density, \sigstar, via:
\begin{equation}
   \left(\dfrac{\sigstar}{\SI{}{\Msun\per\square\parsec}}\right) = \num{4.22e2} \left(\dfrac{I_{\SI{3.6}{\micron}}}{\SI{}{\mega\jansky\per\steradian}}\right) \,,
\end{equation}
which assumes a constant mass-to-light ratio of $0.6\,\SI{}{\Msun}/\SI{}{\Lsun}$ \citep{Meidt2014}.

In App.~\ref{sec:appendix:pde}, we use the stellar mass volume density (\rhostar) to compute the dynamical equilibrium pressure of the ISM in the galaxy disc.
We estimate \rhostar from the stellar mass surface density, adopting the recipes used in \citet{Blitz2006, Leroy2008, Ostriker2010, Sun2020a}:
\begin{align}
    \rhostar = \dfrac{\sigstar}{4H_\star} = \dfrac{\sigstar}{0.54\,R_\star}\,,
\end{align}
where $H_\star$ and $R_\star$ are the scale height and radial scale length of the stellar disc.
Above equation assumes an isothermal density profile along the vertical direction \citep{vanderKruit1988} and a fixed stellar disc flattening ratio of $R_\star/H_\star=7.3$ \citep{Kregel2002}.
We adopt $R_\star=\SI{61.1}{\arcsec}\sim\SI{4.5}{\kilo\parsec}$ from the S$^4$G photometric decomposition analysis \citep{Salo2015}.

\subsection{Dynamical equilibrium pressure}
\label{sec:appendix:pde}
We compute the dynamical equilibrium pressure, or ISM pressure (\PDE) following the prescription by \citet{Sun2020a}.
\PDE describes the pressure regulated by the mass content in the ISM disc and thus provides an important gauge of the local environment of molecular clouds. 
The distribution of stars and gas in a galaxy disc can approximately be described as isothermal fluids in a plane-parallel geometry. 
In this prescription the dynamical equilibrium pressure is composed of a pressure term created by the ISM due to the self-gravity of the ISM disc and a term due to the gravity of the stars (see e.g. \citealp{Spitzer1942}), such that:
\begin{align}
    \PDE = \dfrac{\pi G}{2} \siggas^2 + \siggas \sqrt{2G\rhostar}\, \sigma_{\text{gas,z}} \; ,
    \label{EQU:PDE_large_app}
\end{align}
where we assume a smooth, single-fluid gas disc, and that all gas shares a similar velocity dispersion, so that $\siggas=\sigmol +\sigatom$ is the total gas surface density, \rhostar is the stellar mass volume density near disc midplane and $\sigma_{\text{gas,z}}$ is the velocity dispersion of the gas perpendicular to the disc. 


\citet{Sun2020a} proposed a new formalism which takes into account the self-gravity of the molecular gas at high resolution, i.e. \SI{100}{\parsec} scale. In this work, we adopt their formalism, which is described in the following. To combine the cloud-scale \sigmolcloud data with the large-scale \sigatom and \rhostar, we split \PDE into two parts: the pressure of the molecular gas at cloud-scale, \Pcloud, and the pressure of the smooth extended atomic gas due to the gravity of all gas (atomic and molecular) and the stars, \Patom. The cloud-scale \Pcloud consists of three terms accounting for the self-gravity of the molecular gas, the gravity of larger molecular structures and the gravity of stars:
\begin{align}
    \Pcloud = & \dfrac{3\pi}{8}G\sigmolcloud^2 + \dfrac{\pi}{2}G\sigmolcloud\sigmollarge \nonumber\\ 
    & + \dfrac{3\pi}{4}G\rhostar\sigmolcloud D_{\text{beam}}\;,
    \label{EQU:Pcloud}
\end{align}
where \rhostar is computed as described in App.~\ref{sec:appendix:mstar}. In Equ.~\eqref{EQU:Pcloud}, \sigmolcloud is given at the cloud-scale resolution (here \SI{120}{\parsec}) while \sigmollarge and \rhostar describe the distributions of the molecular gas and the stellar mass density at \SI{260}{\parsec} scale. In typical units, Equ.~\eqref{EQU:Pcloud} reads:
\begin{align}
    &\left(\dfrac{\Pcloud}{\SI{}{\kB\kelvin\per\cubic\centi\metre}}\right) = \num{2.48e5}\left(\dfrac{\sigmolcloud}{\SI{e2}{\Msun\per\square\parsec}}\right)^2 \\
    &\quad + \;\num{3.31e5}\left(\dfrac{\sigmolcloud}{\SI{e2}{\Msun\per\square\parsec}}\right)\left(\dfrac{\sigmollarge}{\SI{e2}{\Msun\per\square\parsec}}\right) \nonumber\\
    &\quad + \;\num{7.01e4}\left(\dfrac{\sigmolcloud}{\SI{e2}{\Msun\per\square\parsec}}\right)\left(\dfrac{\rhostar}{\SI{e-1}{\Msun\per\cubic\parsec}}\right)\left(\dfrac{D_{\text{beam}}}{\SI{150}{\parsec}}\right)\nonumber\;,
\end{align}

The large-scale \Patom includes the self-gravity of the atomic gas and the gravitational interaction of the atomic gas with the (large-scale) molecular gas and the stars:
\begin{align}
    \Patom = \dfrac{\pi G}{2}\sigatom^2 + \pi G \sigatom\sigmollarge + \sigatom\sqrt{2G\rhostar}\,\sigma_{\text{atom}}\;,
    \label{EQU:Patom}
\end{align}
where $\sigma_{\text{atom}}$ is the velocity dispersion of the atomic gas which is fixed at $\sigma_{\text{atom}}=\SI{10}{\kilo\metre\per\second}$. In Equ.~\eqref{EQU:Patom} all quantities are convolved to the large-scale resolution. Converting to typical units we obtain:
\begin{align}
    &\left(\dfrac{\Patom}{\SI{}{\kB\kelvin\per\cubic\centi\metre}}\right) = \num{3.31e5}\left(\dfrac{\sigatom}{\SI{e2}{\Msun\per\square\parsec}}\right)^2 \\
    &\quad + \;\num{6.62e5}\left(\dfrac{\sigatom}{\SI{e2}{\Msun\per\square\parsec}}\right)\left(\dfrac{\sigmollarge}{\SI{e2}{\Msun\per\square\parsec}}\right) \nonumber\\
    &\quad + \;\num{1.02e5}\left(\dfrac{\sigatom}{\SI{e2}{\Msun\per\square\parsec}}\right)\left(\dfrac{\rhostar}{\SI{e-1}{\Msun\per\cubic\parsec}}\right)^{1/2}\nonumber\;,
\end{align}
Finally, we compute the intensity-weighted average, $\langle P_{\rm cloud,\,120\,pc}\rangle_{\rm 260\,pc}$, at the large-scale to combine the molecular gas weight with the large-scale \Patom pixel by pixel:
\begin{align}
    \langle P_{\rm DE,\,120\,pc}\rangle_{\rm 260\,pc} = \langle P_{\rm cloud,\,120\,pc}\rangle_{\rm 260\,pc} + \Patom
\end{align}

\begin{figure*}
\centering
\includegraphics[width=\textwidth]{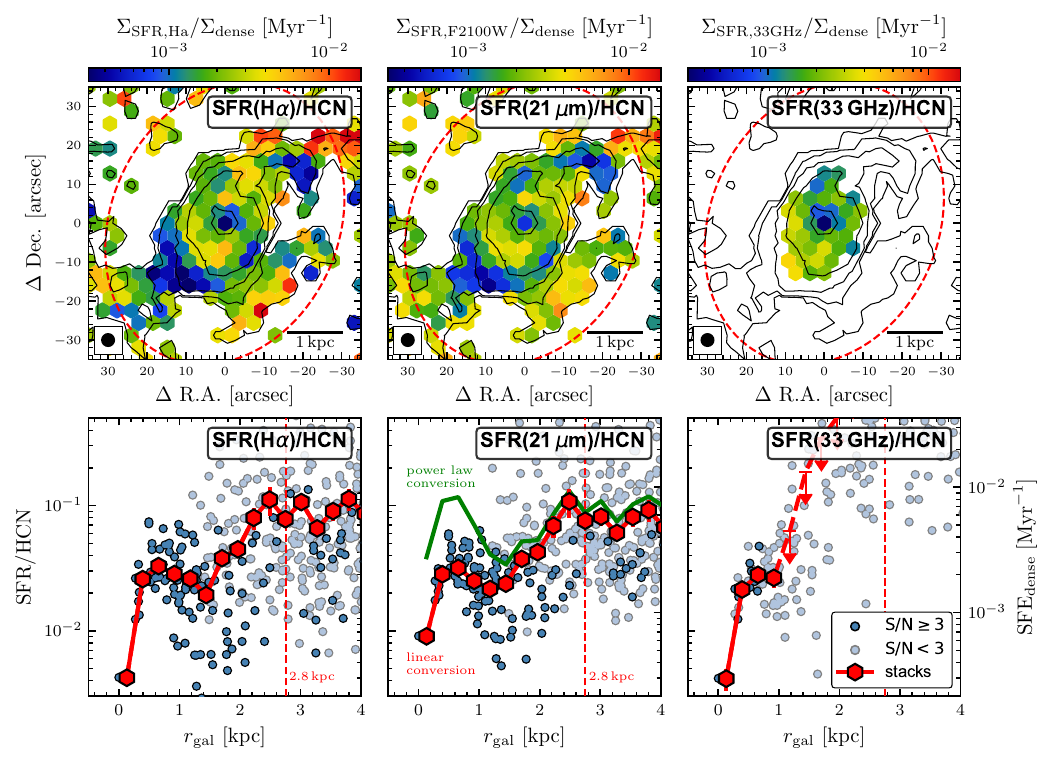}
\caption{Comparison of SFR tracer in dense gas scaling relations in the inner 4$\,$kpc. \textit{Top:} SFR/HCN, tracing dense gas star formation efficiency, using three different SFR tracer from left to right, \halpha from MUSE, \SI{21}{\micron} from JWST and \SI{33}{\giga\hertz} from VLA. Contours show HCN intensities as in Fig.~\ref{fig:hcn_ratio_maps}. The dotted ellipse denotes $r_{\rm gal}=\SI{2.75}{\kilo\parsec}$. \textit{Bottom:} SFR/HCN, matching the respective above panels, against galactocentric radius, $r_{\rm gal}$. Blue points indicate detected ($\mathrm{S/N}\geq3$) and light blue points denote non-significant ($\mathrm{S/N}<3$) data. The red hexagon markers show the spectral stacks taken over all data within the bin. In the middle panel, we show the data obtained from the linear \SI{21}{\micron}-to-SFR conversion (Equ.~\eqref{equ:sfr_21um_lin}) and additionally indicate the mean trend inferred from a power law conversion, i.e. $\mathrm{SFR}\propto L(\mathrm{F2100W})^{1.3}$ \citep{Leroy2023}.}
\label{fig:hcn_ratios_sfr_tracers}
\end{figure*}

\begin{figure*}
\centering
\includegraphics[width=\textwidth]{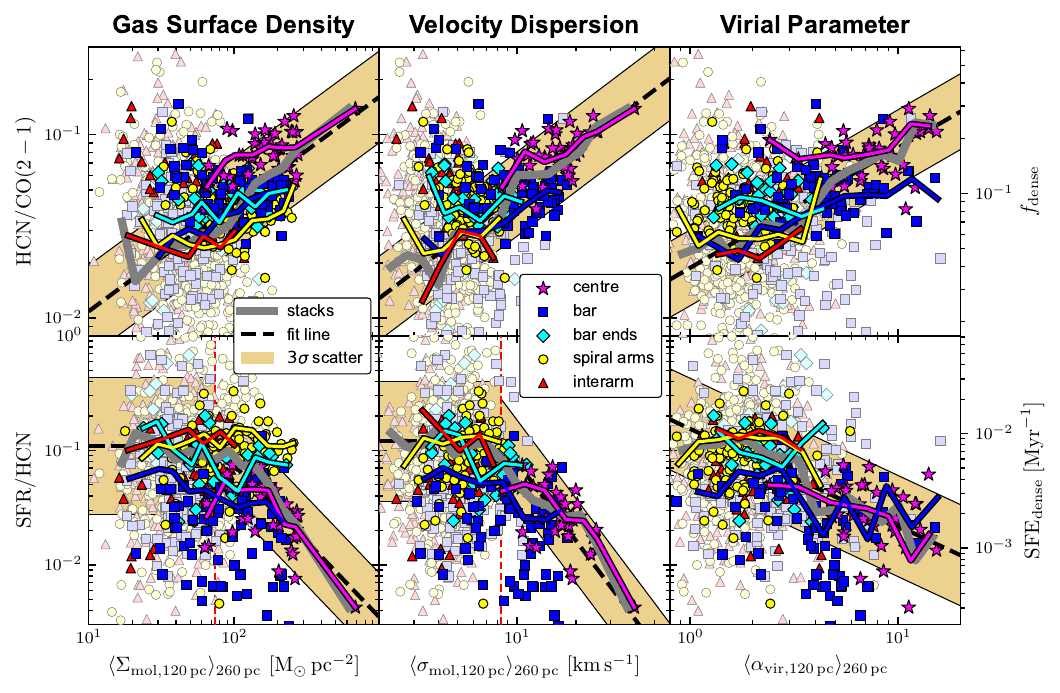}
\caption{HCN spectroscopic ratios against molecular cloud properties. \textit{Top:} HCN/CO and \textit{(bottom)} SFR/HCN measured at \SI{260}{\parsec} scale vs cloud-scale molecular gas properties inferred from \SI{120}{\parsec} scale \cotwo observations. Panels from left to right show molecular cloud surface density (\sigmol), velocity dispersion (\vdis) and virial parameter (\avir) on the $x$-axis, obtained from PHANGS--ALMA as described in Sect.~\ref{sec:appendix:cloud_relations}. Similarly to Fig.~\ref{fig:hcn_ratios_vs_radius_env} and \ref{fig:hcn_ratios_vs_pressure_env}, markers indicate the respective environments and line fits as well as linear regression regimes are determined via MARS. The obtained thresholds are $\sigmol = \SI{74}{\Msun\per\square\parsec}$ for SFR/HCN vs. \sigmol and $\vdis=\SI{8.3}{\km\per\second}$ for SFR/HCN vs. \vdis. The linear regression parameters are listed in Tab.~\ref{tab:fit_parameters_appendix}.}
\label{fig:hcn_ratios_vs_cloud_props}
\end{figure*}

\begin{figure}
\includegraphics[width=\columnwidth]{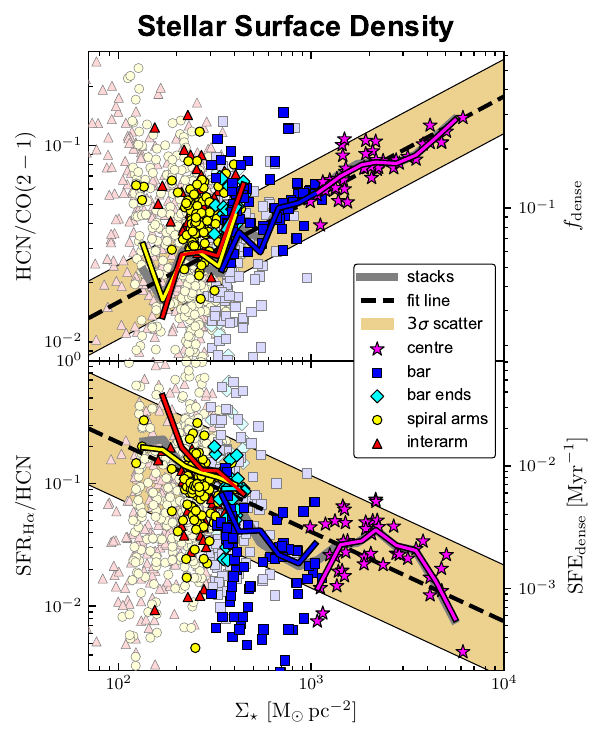}
\caption{HCN spectroscopic ratio vs stellar mass surface density similar to Fig.~\ref{fig:hcn_ratios_vs_radius_env} and \ref{fig:hcn_ratios_vs_pressure_env}, but using \sigstar on the $x$-axis. Stellar mass is traced via the dust-attenuation corrected \SI{3.6}{\micron} emission. The line fit parameters are listed in Tab.~\ref{tab:fit_parameters_appendix}.}
\label{fig:hcn_ratios_vs_stars}
\end{figure}

\begin{figure}
\centering
\includegraphics{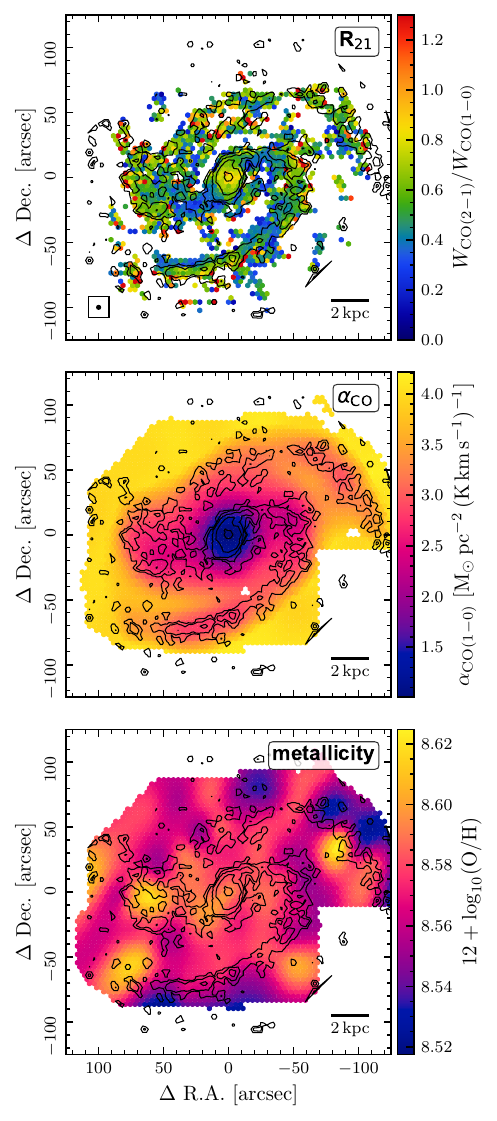}
\caption{Conversion factor maps. \textit{Top:} \cotwo-to-\coone line ratio ($R_{21}$) map computed from \cotwo observations (PHANGS--ALMA) and \coone observations (ALMA science verification program) at a common \ang{;;4} resolution. \textit{Middle:} CO-to-\htwo conversion factor (\aCO) following the prescription from \citet{Bolatto2013} (Equ.~\ref{equ:aco}) using metallicities (\textit{bottom panel}) from PHANGS--MUSE optical recombination lines observations \citep{Williams2022}.}
\label{fig:maps_conversion_factors}
\end{figure}

\begin{figure}
\centering
\includegraphics{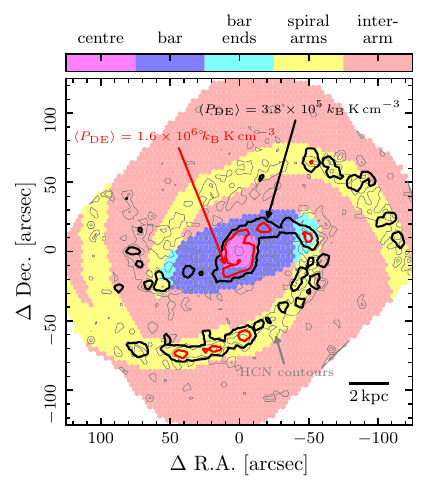}
\caption{Pressure threshold contours. The map shows the same morphological environment masks as in Fig.~\ref{fig:hcn_ratio_maps}. Overlaid are HCN contours in grey and \PDEavg contours at $\PDEavg=\SI{3.8e5}{\kB\kelvin\per\cubic\centi\metre}$ (black) and at $\PDEavg=\SI{1.6e6}{\kB\kelvin\per\cubic\centi\metre}$ (red). The pressure contours represent the threshold values in the pressure relations discussed in Sect.~\ref{sec:results:pressure}.}
\label{fig:maps_pressure_contours}
\end{figure}

\begin{figure}
\centering
\includegraphics{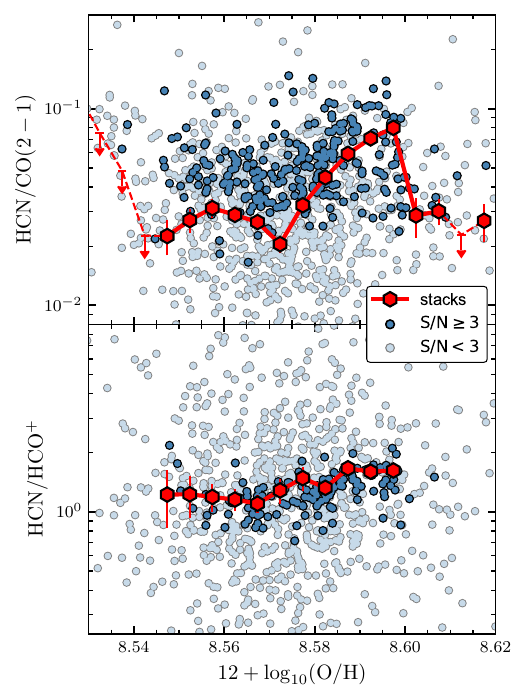}
\caption{Line ratio trends with metallicity. \textit{Top:} HCN/\cotwo against metallicity \textit{Bottom:} HCN/HCO$^+$ against metallicity. Dark markers indicate significant data ($\rm{S/N}\geq 3$) and light markers show non-detections ($\rm{S/N}< 3$). The red hexagon markers show the average trends over all data obtained via spectral stacking. We note that the HCO$^+$ data will be separately published and studied in more detail along with other dense gas tracers in Neumann et al. in prep. Here, we only show the HCN/HCO$^+$ variation with metallicity to highlight the flat trends hence supporting HCN as a tracer of density across the full molecular gas disc of \gal.}
\label{fig:metallicity_trends}
\end{figure}

\begin{figure*}
\centering
\includegraphics[width=\textwidth]{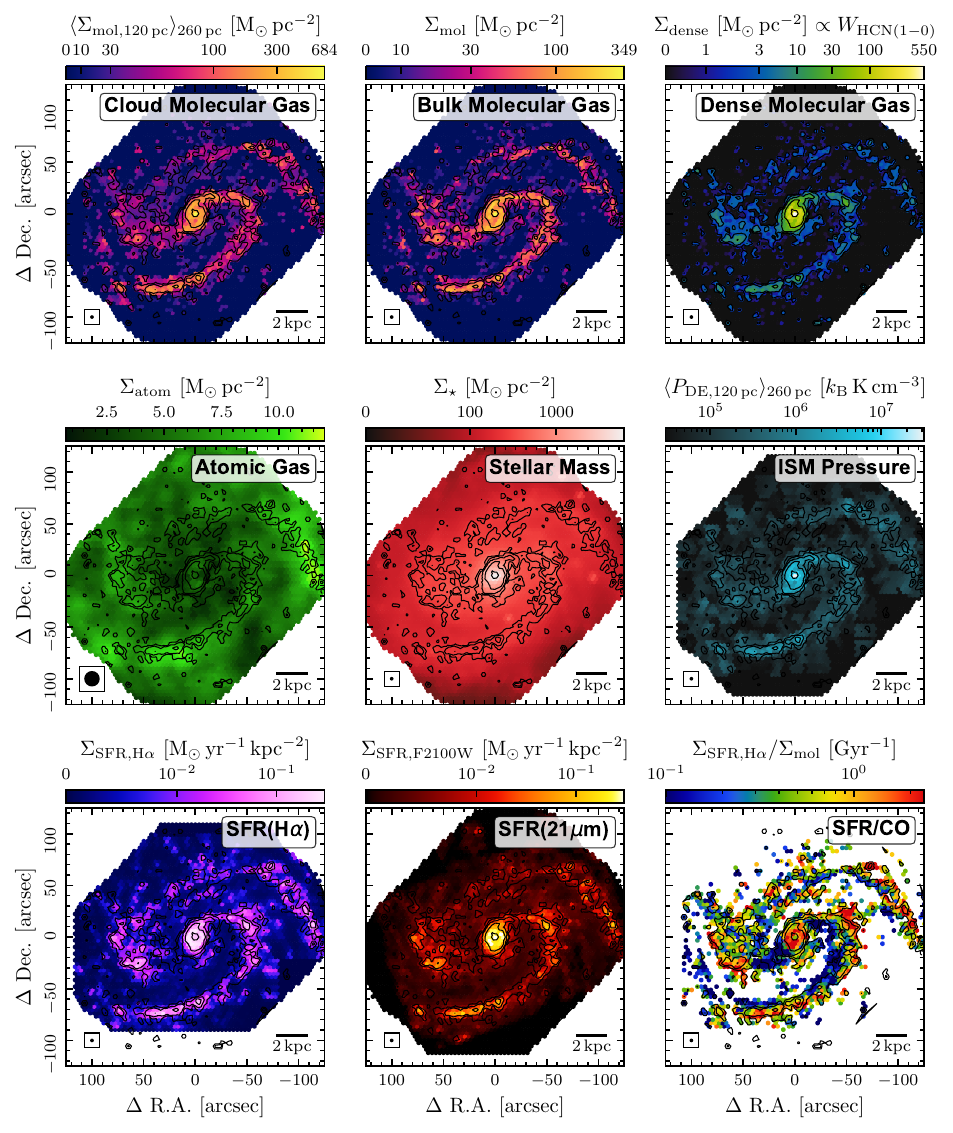}
\caption{Data product maps compilation. All maps are convolved to a common resolution of \SI{260}{\parsec}, given by the native resolution of the HCN data, and sampled to a common, hexagonal pixel grid at beam size spacing.}
\label{fig:data_products_maps}
\end{figure*}

\begin{figure*}
\centering
\includegraphics[width=\textwidth]{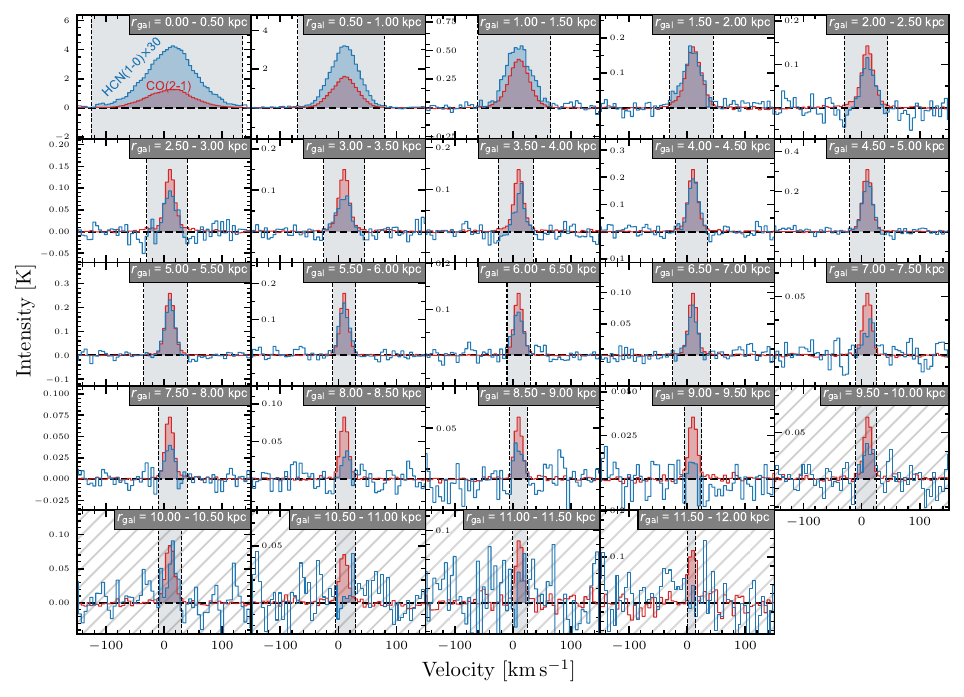}
\caption{Similar to to Fig.~\ref{fig:stacks_envir}, but using the galactocentric radius, \rgal, as the stacking quantity with \SI{0.5}{\kilo\parsec} (twice the beam size) bin widths. The panels with hatched background denote radial bins that are not completely covered by the field-of-view of the observations and thus not considered for the radial fit in Fig.~\ref{fig:hcn_ratios_vs_radius_and_pressure}.}
\label{fig:stacks_radius}
\end{figure*}

\begin{figure*}
\centering
\includegraphics[width=\textwidth]{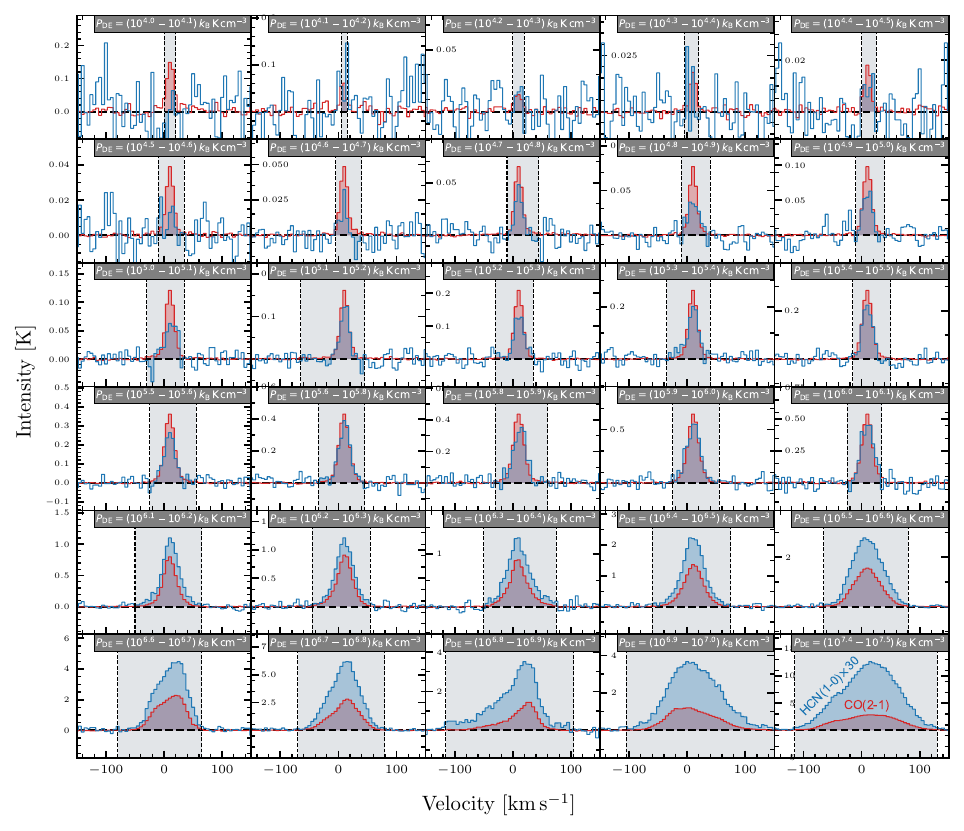}
\caption{Similar to to Fig.~\ref{fig:stacks_envir}, but using the dynamical equilibrium pressure, \PDEavg, as the stacking quantity with \SI{0.1}{\dex} bin widths. Note that the bins from $\PDEavg = 10^{7.0}$ to $10^{7.4}\,\SI{}{\kB\kelvin\per\cubic\centi\metre}$ do not contain any spectra and are therefore not shown.}
\label{fig:stacks_pressure}
\end{figure*}

\begin{figure*}
\centering
\includegraphics[width=\textwidth]{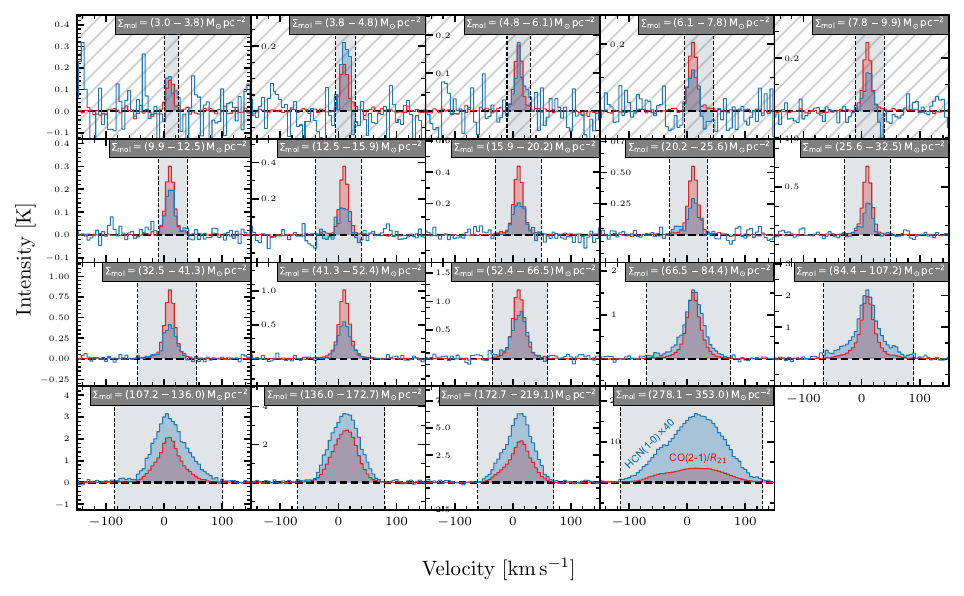}
\caption{Similar to to Fig.~\ref{fig:stacks_envir}, but using the molecular gas surface density, \sigmol, as the stacking quantity with 20 bins from $\sigmol=3-\SI{353}{\Msun\per\square\parsec}$. Note that the bin at $\sigmol=173-\SI{219}{\Msun\per\square\parsec}$ does not contain any spectra and is therefore not shown. Bins with $\sigmol < \SI{10}{\Msun\per\square\parsec}$ are indicated with hatched backgrounds and are not used for the line fit in Fig.~\ref{fig:hcn_co_scaling}.}
\label{fig:stacks_sigmol}
\end{figure*}


\begin{table*}
    \begin{center}
    \caption{HCN/CO($2-1$) and SFR/HCN correlations with \sigstar, \sigmolavg, \vdisavg, \aviravg.}
    \label{tab:fit_parameters_appendix}
    \resizebox{\textwidth}{!}{
    \begin{tabular}{ccccccc}
    \hline \hline
    x-axis & y-axis & Regime & Slope (stacks) & Slope (los) & Corr. ($p$) & Scatter \\ 
    \hline
    \multirow{3}{*}{\sigmolavg} & HCN/CO(2$-$1) & ~ &  0.58 & 0.40 (0.04) & 0.43 (0.0) & 0.25 \\
     & \multirow{2}{*}{SFR/HCN} & $\leq\SI{74}{\Msun\per\square\parsec}$ & 0.00 & -0.75 (0.13) & -0.30 (0.0) & 0.60 \\
     &  & $>\SI{74}{\Msun\per\square\parsec}$ & -1.33 & -2.30 (0.45) & -0.53 (0.0) & 0.39 \\ 
    \hline
    \multirow{3}{*}{\vdisavg} & HCN/CO(2$-$1) & ~ & 0.81 & 0.51 (0.04) & 0.57 (0.0) & 0.23 \\
     & \multirow{2}{*}{SFR/HCN} & $\leq\SI{8.3}{\kilo\metre\per\second}$ & 0.00 & -1.05 (0.26) & -0.27 (0.0) & 0.52 \\
     &  & $>\SI{8.3}{\kilo\metre\per\second}$ & -2.29 & -1.58 (0.16) & -0.60 (0.0) & 0.37 \\ 
    \hline
    \multirow{2}{*}{\aviravg} & HCN/CO(2$-$1) & ~ & 0.65 & 0.46 (0.03) & 0.52 (0.0) & 0.21 \\
     & SFR/HCN & ~ & -0.84 & -1.11 (0.07) & -0.55 (0.0) & 0.44 \\ \hline
    \multirow{2}{*}{\sigstar} & HCN/CO(2$-$1) & ~ & 0.52 & 0.44 (0.03) & 0.57 (0.0) & 0.19 \\
     & SFR/HCN & ~ & -0.73 & -1.02 (0.06) & -0.50 (0.0) & 0.46 \\ 
    \hline\hline
    \end{tabular}
    }
    \end{center}
    {\bf Notes} -- Linear regression parameters for the respective relations and $x$-axis regimes presented in Fig.~\ref{fig:hcn_ratios_vs_stars} and \ref{fig:hcn_ratios_vs_cloud_props} analogous to Tab.~\ref{tab:fit_parameters}. The slopes of the binned relations (column 4) are either determined by \texttt{MARS} if two distinct regimes with different linear regression behaviour have been found, or computed with \texttt{LinMix} if the relation is well described with a single line according to the \texttt{MARS} algorithm.
\end{table*}

\section{Spectral stacking}
\label{sec:appendix:stacking}
We compute spectral stacks using the \texttt{python} package \texttt{PyStacker}\footnote{\url{https://github.com/PhangsTeam/PyStacker}} presented in \citet{Neumann2023b}.
\texttt{PyStacker} uses a high-significance prior (here \cotwo) to determine the velocity field of the line emission.
It then uses this velocity information to shift the spectra of various line to correct for the Doppler shift, assuming all lines share the same velocity field.
Afterwards, we average the shuffled spectra over larger regions usually resulting in higher-significance detections.
Here, we stack the emission of \cotwo (from PHANGS--ALMA) and \hcnone (this work) via morphological environment (Fig.~\ref{fig:stacks_envir}), galactocentric radius, \rgal (Fig.~\ref{fig:stacks_radius}), dynamical equilibrium pressure, \PDEavg (Fig.~\ref{fig:stacks_pressure}), molecular gas surface density (Fig.~\ref{fig:stacks_sigmol}), \sigmol, stellar mass surface density, \sigstar, and molecular cloud properties, \sigmolavg, \vdisavg, \aviravg.
We use these averaged spectra to compute integrated intensities, where the velocity-integration window is inferred based on the average \cotwo spectrum using the same masking method as described in Sect.~\ref{sec:methods:mom0_maps}.
We note that the spectral stacks agree within $\pm \SI{10}{\percent}$ with binned integrated intensities computed within the same stacking bins if \cotwo is used as a prior to define the velocity-integration mask.
\section{SFR tracers}
\label{sec:appendix:sfr_tracers}

\subsection{SFR -- \SI{21}{\micron}}
\label{sec:appendix:sfr_21mu}



We use \SI{21}{\micron} (F2100W) emission from recent JWST--MIRI \citep{jwst-mission, miri-paper} observations as a another probe of SFR, in addition to \halpha from PHANGS--MUSE (Sect.~\ref{sec:methods:sfr_halpha}) and \SI{33}{\giga\hertz} from VLA (Sect.~\ref{sec:appendix:sfr_33ghz}).
These data are part of the ``PHANGS--JWST Treasury Program'' \citep{Lee2023} and have been reduced via the PHANGS--JWST data reduction pipeline (Williams et al. in prep.).
\gal was observed by JWST in June 2023 and we use version 0.9 of the PHANGS--JWST data reduction.

Physically, the strong radiation field from young, massive stars heat up the surrounding dust, which re-emits at infrared wavelength probed by F2100W.
\SI{21}{\micron} point sources correlate well with $\text{H}\,{\tiny\text{II}}$ regions \citep{Hassani2023} and the F2100W intensity correlates well, though non-linearly, with extinction-corrected \halpha intensity \citep[][]{Leroy2023, Belfiore2023}. 
However, the F2100W also captures stochastically heated emission from small dust grains that can trace the ISM. 
It may thus be both more robust to extinction than \halpha and more subject to contamination by diffuse ISM emission.

To infer \sigsfr from F2100, we use the empirical relation from \citet{Leroy2023} (their equation 5), which re-scales the \SI{21}{\micron} flux into a \SI{24}{\micron} flux using $R_{\SI{21}{\micron}/\SI{24}{\micron}}=0.80$ and then converts to SFR via a linear conversion \citep[e.g. following][]{Kennicutt2012}, such that the SFR surface density is given by:
\begin{equation}
    \left(\dfrac{\Sigma_{\rm SFR, F2100W}}{\SI{}{\Msun\per\year\per\square\kilo\parsec}}\right) = \num{3.7e-3} \left( \dfrac{I_{\nu, \rm{F2100W}}}{\SI{}{\erg\per\second\per\square\centi\metre\per\steradian}} \right)\cos(i) \;,
    \label{equ:sfr_21um_lin}
\end{equation}
Thus, we adopt a linear relation between F2100W dust emission and the SFR.
We also contrast this prescription to a power law relation based on \citet{Leroy2023} which leads to up to a factor of three higher values in the central $\sim\SI{2}{\kilo\parsec}$ of \gal.

\subsection{SFR -- \SI{33}{\giga\hertz}}
\label{sec:appendix:sfr_33ghz}

We note that above SFR tracers might lead to significantly discrepant results in galaxy centres, where optical recombination lines can become too extinct to recover robust Balmer decrement corrections and \SI{21}{\micron} emission might be systematically contaminated by stochastically heated dust grain.
Therefore, we use free-free \SI{33}{\giga\hertz} emission from as an additional SFR tracer in the centre of \gal.
The data are coming from Very Large Array (VLA) observations of a large sample of galaxies, including \gal, at \SIrange{3}{33}{\giga\hertz} at $\sim\ang{;;2}$ resolution \citep{Linden2020}.

At high radio frequencies the ionising flux of young massive stars is directly proportional to the thermal spectral luminosity. 
This allows us to trace the SFR via the thermal part of the \SI{33}{\giga\hertz} flux measured by the VLA following the prescription in \citet{Murphy2012} (their equation 6):
\begin{align}
    \left(\dfrac{\Sigma_{\rm SFR,\SI{33}{\giga\hertz}}}{\SI{}{\Msun\per\year\per\square\kilo\parsec}}\right) = & \, \num{5.5e16} \, \left( \dfrac{T_{\rm e}}{\SI{e4}{\kelvin}} \right)^{-0.45} \left( \dfrac{\nu}{\SI{}{\giga\hertz}} \right)^{0.1} \, f_{\rm thermal} \nonumber\\ 
    & \times \left( \dfrac{I_{\nu,\SI{33}{\giga\hertz}}}{\SI{}{\erg\per\second\per\square\centi\metre\per\steradian}} \right) \,\cos(i) \;,
    \label{equ:sfr_33ghz}
\end{align}
where the thermal fraction ($f_{\rm thermal}$) values are taken from \citet{Linden2020} (their table 4; $\sim\SI{200}{\parsec}$ apertures) and we adopt an electron temperature of $T_{\rm e}=\SI{e4}{\kelvin}$ and $\nu=\SI{33}{\giga\hertz}$.

\subsection{SFR tracer comparison}
\label{sec:appendix:sfr_comparison}

Throughout this work, we have used Balmer decrement-corrected \halpha emission as a tracer of SFR (Sect.~\ref{sec:methods:sfr_halpha}).
Despite being overall a robust tracer of SFR, when corrected for dust attenuation, in the centres of galaxies, \halpha might miss some of the SFR-related \halpha emission due to extremely high attenuation by denser dust thus biasing the SFR estimate high.
Also, in extreme environments, like the centre, \halpha emission might be associated with other processes than young stars and thus bias the SFR high.
Therefore, the question is: ``How robust are our findings against the choice of the SFR tracer?''

We use JWST F2100W emission to trace SFR via the hot dust emission that is less affected by attenuation effects in the centre of the galaxy as long as these galaxy centres contain sufficient amounts of dust (Sect.~\ref{sec:appendix:sfr_21mu}).
Using the F2100W-inferred SFR (instead of \halpha) leads to similar SFR/HCN values across the disc and in the low-pressure regime for both prescriptions (linear and power-law conversion) adopted here (see Fig.~\ref{fig:hcn_ratios_sfr_tracers}.
However, the two calibrations differ by up to a factor of three in the central kpc of the galaxy.
The change is so severe that using the power-law calibration yields similar SFR$_{\rm F2100W}$/HCN values in the centre (median $10^{-1.05}$) as in bar ends (median $10^{-0.96}$), spiral arms (median $10^{-1.06}$) and interarm (median $10^{-1.19}$), and statistically the same SFR$_{\rm F2100W}$/HCN distribution in the centre compared to the aforementioned environments.

In addition, we compare the two SFR tracers (\halpha and F2100W) with the \SI{33}{\giga\hertz}-inferred values, being the most robust tracer of the SFR in the centre of \gal (Sect.~\ref{sec:appendix:sfr_33ghz}).
We find that \halpha (decrement-corrected) and \SI{33}{\giga\hertz} yield very consistent results across the whole central \SI{1}{\kilo\parsec} probed by \SI{33}{\giga\hertz} emission, agreeing within \SI{20}{\percent}.
F2100W (linear conversion) leads to a factor of two higher values in the central \SI{260}{\parsec}, but otherwise consistent values.
Therefore, we conclude that for \gal Balmer decrement-corrected \halpha emission is an excellent tracer of the SFR, even in the centre of the galaxy, where increased dust attenuation could have depreciated \halpha as a robust SFR tracer.
Moreover, these results suggest that F2100W is proportional to the SFR across the full disc of \gal.

\section{Scaling relations}
\label{sec:appendix:scaling_relations}

\subsection{Molecular cloud relations}
\label{sec:appendix:cloud_relations}

In Fig.~\ref{fig:hcn_ratios_vs_cloud_props}, we present the relations between HCN/CO and SFR/HCN, respectively with the properties of molecular gas at GMC scales, i.e. here, at \SI{120}{\parsec}.
The cloud-scale molecular gas properties are computed following the prescriptions of \citet{Sun2018} and \citet{Neumann2023a}, but adopting the varying \aCO and $R_{21}$ conversions introduced in Sect.~\ref{sec:methods:sigmol_varying_aco}.
We compute the molecular gas surface density from the \cotwo integrated intensity (\intCO):
\begin{align}
    \sigmol = \aCO \, R_{21}^{-1} \, \intCO \,.
\end{align}
The molecular gas velocity dispersion is obtained from the \cotwo equivalent line width and corrected for the velocity channel-to-channel correlation:
\begin{align}
    & \sigma_{\rm measured} = \dfrac{\intCO}{\sqrt{2\pi}T_{\rm peak}} \,,\\
    & \vdis = \sqrt{\sigma_{\rm measured}^2 + \sigma_{\rm response}^2} \,,
\end{align}
where $\sigma_{\rm measured}$ assumes a Gaussian line profile and $\sigma_{\rm response}$ takes into account the instrument channel width and channel-to-channel correlation.
To estimate the virial parameter (\avir), we assume spherically symmetric clouds with a given density profile $\rho\propto r^{-1}$, so that \avir can be computed from the \cotwo data using \sigmol and \vdis:
\begin{align}
    \avir = \dfrac{9\,\ln{2}}{\pi\,G} \dfrac{\vdis^2}{\sigmol\,D_{\rm beam}} \propto \dfrac{\vdis^2}{\sigmol}\,,
\end{align}
where $G$ is the gravitational constant.
These molecular gas properties are computed at $D_{\rm beam} =\SI{120}{\parsec}$ scale and converted to the HCN resolution, i.e. \SI{260}{\parsec}, via a \sigmol-weighted average, similar to \PDE as described in Sect.~\ref{sec:methods:pde}, hence the notation $\langle X_{\SI{120}{\parsec}}\rangle_{\SI{260}{\parsec}}$, where $X$ is the quantity to by averaged.

Similar to the relations with radius (Sect.~\ref{sec:results:radius}) and pressure (Sect.~\ref{sec:results:pressure}), we apply the MARS linear regression tool (Sect.~\ref{sec:methods:fitting}) to the cloud property relations shown in Fig.~\ref{fig:hcn_ratios_vs_cloud_props}.
We find that HCN/CO correlated positively and with \sigmol, \vdis and \avir and is well-described by a single power law over the whole data range that is probed.
Similarly, we observe a negative correlation between SFR/HCN and \sigmol, \vdis, \avir.
However, for SFR/HCN vs \sigmol and \vdis we find a change in the relation (as determined via MARS) at $\sigmol=\SI{74}{\Msun\per\square\parsec}$ and $\vdis=\SI{8.3}{\km\per\second}$, such that SFR/HCN is roughly constant at low \sigmol (\vdis) and strongly decreasing with increasing \sigmol (\vdis) at high \sigmol (\vdis).

\subsection{Stellar mass relations}
\label{sec:appendix:mstar_relations}

Fig.~\ref{fig:hcn_ratios_vs_stars} shows the HCN/CO and SFR/HCN scaling relation with the stellar mass surface density, \sigstar.
We compute \sigstar from the \textit{Spitzer} \SI{3.6}{\micron} image as described in Sect.~\ref{sec:appendix:mstar}.
Following the same methodology as in Sect.~\ref{sec:results:radius}, ~\ref{sec:results:pressure}, ~\ref{sec:appendix:cloud_relations}, we determine the power law behaviour of the mean trend finding strong correlations between HCN/CO (positive correlation) and SFR/HCN (negative correlation) with \sigstar.
For HCN/CO vs. \sigstar, we observe a tighter relation (\SI{0.19}{\dex} scatter) that is consistent over all environments and the whole probed data range, spanning two orders of magnitude in stellar mass surface density ($\SI{1e2}{\Msun\per\square\parsec}$ to $\SI{1e4}{\Msun\per\square\parsec}$).
The found relation shows that \sigstar is a good predictor of HCN/CO over the whole disc of a nearby galaxy at sub-kpc scales, though most of the dynamic range of \sigstar is covered by only two regions, the centre and the bar.
The relation between SFR/HCN and \sigstar shows much higher scatter (\SI{0.46}{\dex}) with the bar region being offset to the main relation by \SI{0.2}{\dex}, stressing the strongly suppressed star formation efficiency in the bar (Sect.~\ref{sec:discussion:bar}).
Combined with the results presented in Sect.~\ref{sec:appendix:cloud_relations}, the found relations suggest that the threshold behaviour in the SFR/HCN vs \PDE relation (Sect.~\ref{sec:results:pressure}) is caused by molecular gas cloud-scale physics rather than larger scale environment.

\section{Additional figures}

Fig.~\ref{fig:maps_conversion_factors} shows the \cotwo-to-\coone line ratio ($R_{\rm 21}$) computed from the $\SI{4}{\arcsec}\sim\SI{300}{\parsec}$ resolution \coone data from \citet{Pan2017} and homogenised \cotwo observations form PHANGS--ALMA \citep{Leroy2021b}.
We use the measured \SI{300}{\parsec}-scale $R_{\rm 21}$ map to infer the cloud-scale (\SI{120}{\parsec}) \coone line intensities from the observed \cotwo line intensities, which enter the estimation of the dynamical equilibrium pressure (Sect.~\ref{sec:methods:pde}).
In Sect.~\ref{sec:discussion:hcn_co}, we use the $R_{\rm 21}$ map to convert the \SI{260}{\parsec}-scale $W_\cotwo$ into $W_\coone$ to compare the HCN-to-\coone vs. \sigmol scaling relation with literature findings.

In Fig.~\ref{fig:maps_pressure_contours}, we show a map of the morphological environments overlaid with HCN contours similar to Fig.~\ref{fig:hcn_ratio_maps}, right panel.
In addition, we indicate two loci of ISM pressure, \PDE, matching the pressure thresholds inferred for the HCN/CO and SFR/HCN vs \PDE scaling relations (Sect.~\ref{sec:results:pressure}).

Fig.~\ref{fig:metallicity_trends} presents HCN/CO and HCN/HCO$^+$ line ratio trends with metallicity. Detected sigh line measurements of HCN/CO show no correlation with metallcity ($\rho_{\rm Pearson}=0.12$). 
The stacked average HCN/CO increases with metallicity, suggesting that the HCN abundance is more strongly depending on metallicity than CO, which is expected if nitrogen decreases more sharply with metallicity than oxygen \citep{Braine2017, Braine2023}.
In this scenario we would expect a systematic increase of the HCN/HCO$^+$ line ratio with metallicity.
However, we find only a weak correlation between HCN/HCO$^+$ and metallicity ($\rho=0.35$), indicating that metallicity effect play only a minor role across \gal, potentially due to the small dynamic range in metallicity, spanning less than \SI{0.1}{\dex} (8.54 to 8.62).

In Fig.~\ref{fig:data_products_maps}, we display maps of various data products surface density maps of the atomic gas, (dense) molecular gas and stellar mas, ISM pressure and SFR inferred from \halpha and \SI{21}{\micron}, respectively.
In addition, we show the star formation efficiency of the molecular gas, SFR/CO.

\end{appendix}

\end{document}